\documentclass[titlepage=firstiscover]{scrbook}
\usepackage[utf8]{inputenc}
\usepackage{epigraph}
\usepackage{graphicx}
\usepackage{hyperref}
\usepackage[square, numbers, sectionbib]{natbib}
\usepackage{chapterbib}
\usepackage[nottoc,notlot,notlof]{tocbibind}
\usepackage{physics}
\usepackage{pdfpages} 

\usepackage{amsmath}
\usepackage{amssymb}

\hypersetup{
    colorlinks = true,
    linktocpage = true,
    linkcolor = blue
}

\title{Generación de entrelazamiento y simulación relativista con osciladores paramétricos en cQED \\ \vspace{.75cm} Entanglement generation and relativistic simulation with cQED parametric oscillators}
\author{Andrés Agustí Casado}
\date{}
\titlehead{\centering Submitted to \textit{Universidad Complutense de Madrid, Facultad de Ciencias Físicas} for the degree of Doctor in Physics}
\publishers{
    \includegraphics[]{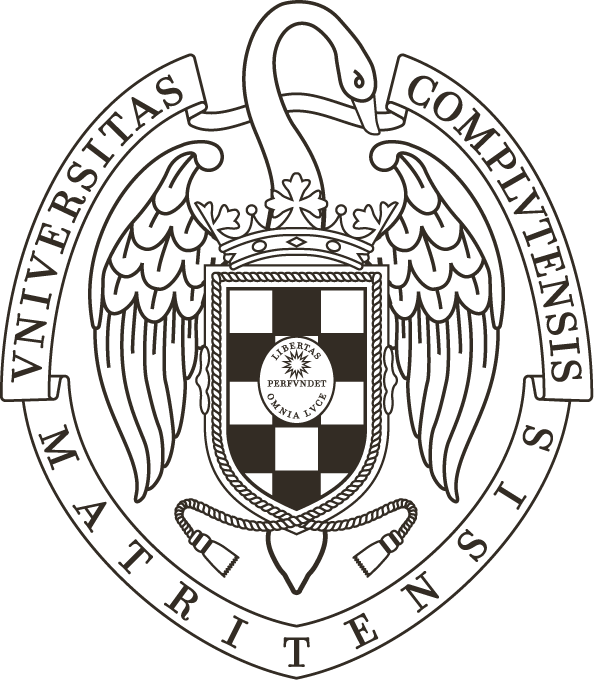} \\
    \vspace{.75cm}
    Written under the supervision of \\ Dr. Carlos Sabín Lestayo \\
    at Instituto de Física Fundamental, CSIC
    }

\begin{document}
    \maketitle

    \frontmatter

    \includepdf[pages=-]{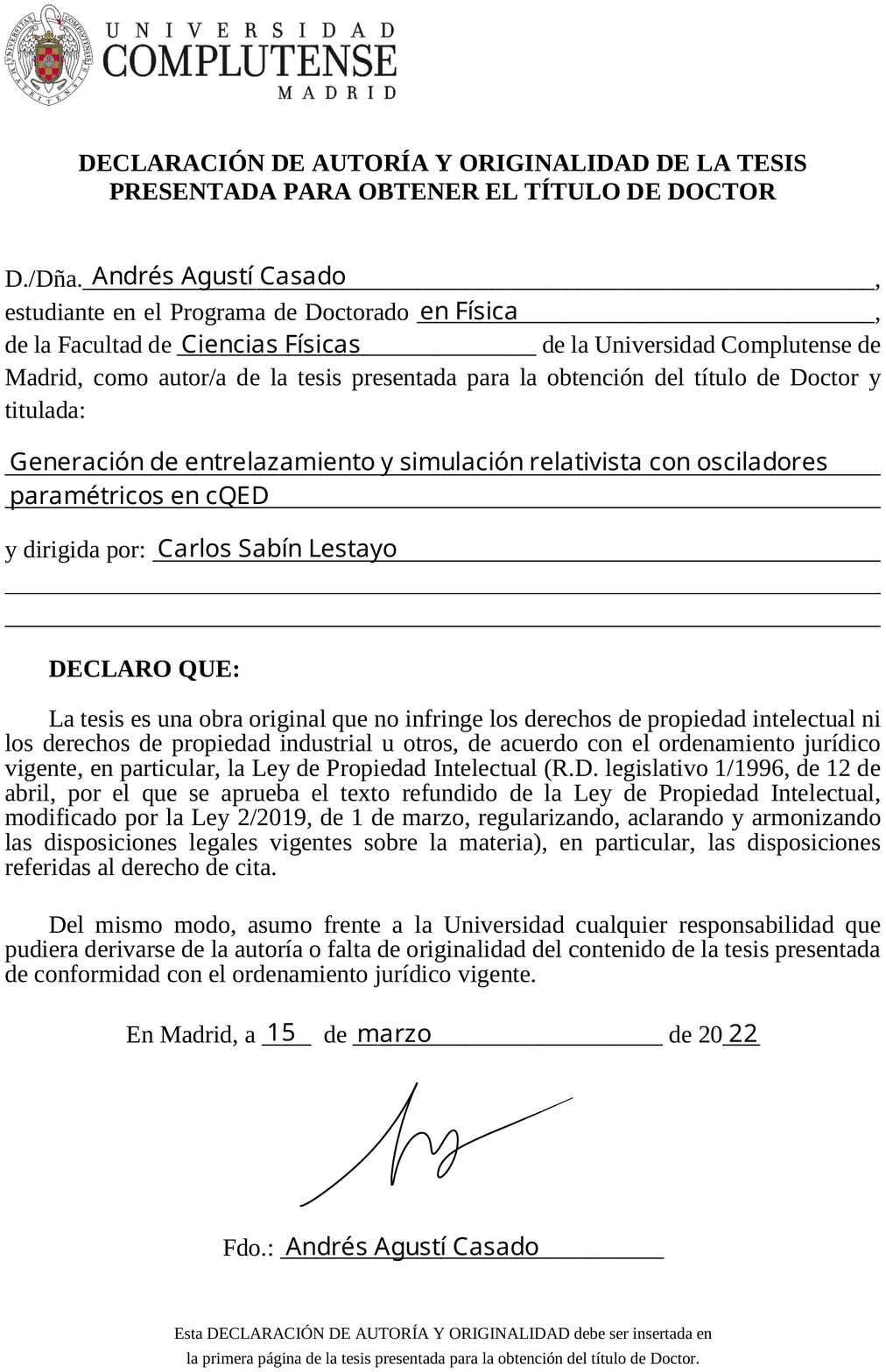}
	\tableofcontents
	\chapter{Resumen}
	\label{sec:sintesis}

    Estamos viviendo una época en la que nuestra comprensión de la mecánica cuántica está  avanzando a un ritmo excitante. Desde su concepción, siempre se le ha considerado el cimiento de nuestro entendimiento sobre una gran variedad de fenómenos físicos. Sin embargo, su papel solía consistir en ser un primer paso hacia la creación de teorías efectivas sobre, por ejemplo, las propiedades de semiconductores o la estadística de procesos que impliquen partículas fundamentales. Pero hoy en día, podemos explorar sus características más íntimas y sorprendentes con la más exquisita de las precisiones. O dicho de otra forma, se aproxima el tiempo en el que sea una tecnología.

    El desarrollo de las tecnologías cuánticas está claramente motivado por sus aplicaciones. Las peculiaridades de la mecánica cuántica se han convertido en oportunidades para el almacenamiento, la transferencia y el procesamiento de información de formas imposibles para las ciencias clásicas. Por lo tanto, el diseño de computadores  cuánticos capaces de otorgarnos esas ventajas es una tarea central para la investigación actual.

    Esta intersección, entre curiosidad científica y posibilidad tecnológica, es donde esta tesis se desarrolla. Se prestará atención al estado del arte en tecnologías cuánticas, en particular a la electrodinámica cuántica de circuitos (cQED, por sus siglas en inglés). Este término engloba a cualquier circuito diseñado con material superconductor, normalmente incluyendo versátiles uniones Josephson. Se modelizarán circuitos existentes o se propondrán nuevos diseños que puedan responder preguntas actuales sobre mecánica cuántica. Esta tesis explora dos de esas preguntas. 

    Por un lado, se investiga la generación y detección de nuevos estados entrelazados no gaussianos de la radiación microondas. Estos estados son producidos en un nuevo oscilador paramétrico, construido recientemente en el ámbito de cQED, capaz de convertir un tono microondas en tres tonos diferentes simultáneamente. Esos nuevos tres fotones comparten entre sus magnitudes correlaciones cuánticas, en particular entrelazamiento genuino, que es considerado el recurso que permite el procesamiento cuántico de la información imposible a las ciencias clásicas. En el texto este entrelazamiento es denominado como no gaussiano debido a que no se manifiesta en segundos momentos estadísticos y proponemos un criterio simple y práctico para la creación de testigos capaces de detectarlo: deben construirse con momentos estadísticos superiores, que cambien en el tiempo. Adicionalmente, se exploran las implicaciones teóricas que puede tener este criterio y se encuentran conexiones con otras clases de entrelazamiento, como la paradigmática inequivalencia entre los estados GHZ y W de tres qubits.

    Por otro lado, se explora otra de las posibles aplicaciones de las tecnologías cuánticas: la simulación de sistemas cuánticos. En la literatura anterior a la presente tesis abundan los circuitos que imitan sistemas en los que se deben considerar simultáneamente fenómenos cuánticos y relativistas, como pueden ser los efectos Casimir dinámico y Unruh. Estos últimos no han sido observados en sus formulaciones originales sino, alguno, en circuitos que crean fenómenos análogos. En este trabajo se explora la información que puede obtenerse de dichos análogos, proponiendo un circuito que permita explorar la dinámica interna de un espejo experimentando una trayectoria relativista, es decir, un espejo que produce el efecto Casimir dinámico.

	\chapter{Abstract}
	\label{ch:abstract}
	We are living in an age where our understanding of quantum mechanics is increasing at an exciting pace. Since its inception, we have always considered it at the foundation of a great variety of physical phenomena. However, it was often used as a first step in the creation of effective theories about, for instance, properties of semiconductors or statistics of processes involving fundamental particles. But nowadays, we are learning to harness the theory to its full extent. Nowadays, we can explore its most intimate and shocking features with the most exquisite control. It is close to become a technology.

	The development of quantum technologies is clearly motivated by their applications. The oddities of quantum mechanics have become opportunities to store, transfer and process information in ways impossible to the classical sciences. Therefore, the task of designing a quantum computer capable of delivering said advantages is central to the research being done today.

	This interface, between scientific curiosity and technological possibility, is where this thesis stands. Attention will be paid to the current state of the art in quantum technologies, mainly in circuit quantum electrodynamics (cQED). This acronym is an umbrella term that encompass any circuit designed to operate with superconducting material, most often composed of versatile Josephson junctions. Then, we will model existing circuits or propose new designs that may shed light on interesting topics in quantum mechanics. This thesis explores two specific topics:

    On one hand, we study the generation and detection of new entangled non-gaussian states of microwave radiation. These states are produced in a new parametric oscillator, built recently within the field of cQED, capable of down-converting a microwave tone into three different tones at once. These new three photons share among their magnitudes quantum correlations, in particular genuine entanglement. This kind of entanglement is considered one of the resources that allow for the quantum processing of information impossible to the classical sciences. In this text we refer to it as non-Gaussian because of its manifestation on statistical moments higher than covariances and we propose a simple and practical criterion for the design of witnesses capable of detecting it: they must be built from higher statistical moments that change through time. Additionally, we speculate on the theoretical implications of the criterion and find suggestive connections to other entanglement classes, such as the paradigmatic nonequivalent GHZ and W three qubit states.

    On the other hand, we explore one of the possible applications of quantum technologies: the simulation of quantum systems. The literature prior to this thesis showcases multiple examples of superconducting circuits capable of mimicking systems in which one must consider both quantum and relativistic phenomena, such as the dynamical Casimir and Unruh effects. These effects have not been observed in their original setting, and some have only been observed in their analogue circuits. This work explores the information that can be obtained through analog simulation, proposing a circuit capable of featuring the internal dynamics of a mirror experiencing a relativistic trajectory, that is, a mirror producing the dynamical Casimir effect.

	\chapter{List of Publications}
    This thesis presents the results published in the following peer-reviewed articles
    \begin{enumerate}
        \item \textbf{A. A. C.}, C. W. S. Chang, F. Quijandría, G. Johansson, C. M. Wilson and C. Sabín, \href{http://dx.doi.org/10.1103/physrevlett.125.020502}{Physical Review Letters \textbf{125} 020502} (2020)
        \item \textbf{A. A. C.}, L. García Álvarez, E. Solano and C. Sabín, \href{https://doi.org/10.1103/physreva.103.062201}{Physical Review A \textbf{103} 062201} (2021)
        \item \textbf{A. A. C.} and C. Sabín \href{https://link.aps.org/doi/10.1103/PhysRevA.105.022401}{Physical Review A \textbf{105} 022401} (2022).
    \end{enumerate}
    Additionally, during the time the author of this thesis spent within the \textit{Universidad Complutense de Madrid} Doctorate program, another publication took place
    \begin{itemize}
        \item P. Cordero Encinar, \textbf{A. A. C.} and C. Sabín \href{https://link.aps.org/doi/10.1103/PhysRevA.104.052609}{Physical Review A \textbf{104} 052609} (2021).
    \end{itemize}

	\mainmatter
	\chapter{Introduction}
	\label{intro}
    The main purpose of chapter \ref{intro} is to present the reader an overview of the work contained in this thesis. There are three intentions behind such a summary. Firstly, we expect most of our readers to be interested only in pieces of the thesis, perhaps because they suspect our research might be connected to theirs. Therefore, the present overview intends to guide them to the pieces they are interested in, see section \ref{intro-navigation}. Secondly, other group of readers will travel through the whole thing. For them, the overview provides a rationale, a thread of thoughts that encompass this thesis entirely, which is provided in section \ref{intro-rationale}. Thirdly, this work builds upon the research of many others. We consider it is unfair to imagine the readers are familiar with all that previous research. Therefore, this chapter introduces some concepts on multipartite entanglement in section \ref{sec:intro-entanglement}. Additionally, we provide references that we recommend to readers not familiar with circuit quantum electrodynamics (cQED) in section \ref{sec:intro-cQED}
    
    \section{Objectives}
        \label{intro-rationale}
        Quantum theory has been a successful theory for around a century. As time passed, more and more phenomena were satisfactorily predicted, up to the point of considering it the most relevant foundation of our understanding of nature, together with general relativity. However, our attitude towards quantum effects has not always been the same. In its early years, its probabilistic and non-local features collided with our prejudices on the workings of nature. Quantum mechanics was considered a faulty tool that needed improvement, an instrument that needed getting rid of those features. With time we saw that was not the case, those striking features did not imply worrisome relations with consciousness, in the case of the measurement problem, or with faster than light actions, in the case of non-locality. Today, coherence and entanglement are considered the defining properties of quantum theories. Today, we understand the importance of exploring the consequences of uncertainty relations and non-locality.

        It took more than theoretical work to change our opinions in this way. It took experimenting. The history of the XX and XXI centuries spans a chain of experiments that, ranging from the early frequency-discrete radiation of atoms to the surprising teleportation of quantum states, has discovered us increasingly intimate phenomena of quantum theory. The realization of those experiments and many others has granted us technologies that allow for the control and design of physical systems previously ignored. In order to exemplify this paradigm shift in greater detail, consider the text-book postulates of quantum mechanics: they specify what possible states a system may have (rays on any Hilbert space), what possible dynamics can take place (any Hamiltonian and measurements) and how to compose systems together (tensor products of states and operators); but they all rely on external theories to specify the actual content of the Universe. One of those theories is quantum electrodynamics: it tells us what is the particular set of possible states of the electromagnetic field and charged particles, the specific dynamics (the Hamiltonian) that will govern them and so on. Of course, those externals theories are, to our knowledge, correct and valuable all the way to the Standard Model of Fundamental Particles. But in the present day we can put the focus back on quantum mechanics and away from the external theories. Nowadays systems and hamiltonians are engineered from the more basic ingredients that the external theories postulate. 
As such, we design qubits with features of our choosing, instead of letting external theories specify which features physical systems have,
        or we design interactions between particular pairs or triplets of systems instead of figuring out how many nearest neighbours we must consider in our models. In this new paradigm we can explore the consequences of those postulates in a deeper sense, we can \textit{choose} the content of our Universe to take quantum effects to its ultimate consequences.
    
        Having described this mindset, this thesis is about using present-day quantum technologies to focus our theoretical work on quantum phenomena. With that context in mind, we explore systems that may contest our understanding of pieces of the theory. Lastly, we try to close the loop by proposing experiments that may prove or disprove the theory, as well as create new resources for further theoretical and practical developments.

    \section{Navigating this thesis}
        \label{intro-navigation}
        Following the lines of the last paragraph of section \ref{intro-rationale}, we have taken concepts of a very versatile technology, circuit quantum electrodynamics (cQED), and used them to explore two different projects: genuinely tripartite entanglement and analog simulation of quantum-relativistic phenomena. Those two ideas led to three publications around which the chapters to come revolve.
        
        Chapter \ref{chapter1} is about a proposal of an entanglement detection protocol. Said entanglement is thought to be produced by an already engineered three-body interaction in bosonic systems, that we denominate three-mode spontaneous parametric down conversion. Said interaction is known for generating non-gaussian states, so we will denominate its entanglement as non-gaussian throughout this work.
        
        Chapter \ref{chapter2} is about the modification of a common Hamiltonian in cQED and other fields, the Rabi Hamiltonian, so that it may become the platform for the analog simulation of the microscopical basis of the dynamical Casimir effect. Said modification consists in the ability to tune the qubit-field coupling over time, which very much relates to the movement, real or simulated, of the qubit within the field modes. Then, we identify the qubit movement as if it were the movement of a mirror, providing the field with energy so that vacuum might be populated by photons.
        
        Chapter \ref{chapter3} is about non-gaussian entanglement. In particular, we provide a precise definition for said entanglement and generalise the application of the concepts to systems beyond the bosonic modes of chapter \ref{chapter1}. We find connections to the \textit{GHZ} and \textit{W} classes of pure 3 qubit states and we strengthen the results of the previous chapter.
        
        Finally, we point out that chapters \ref{chapter1} to \ref{chapter3} share a similar structure. They revolve around a publication in a peer-reviewed journal. Therefore, we take the opportunity of this thesis to broaden the target audience of our work, which implies that those chapters contain accessible introductions to those publications, as well as providing greater detail in the derivation of their results. Additionally, we provide some materials that did not make it into the publications, as well as a discussion of the different connections between them. Note that throughout the text, $\hbar = 1$. Furthermore, we commit an abuse of notation and frequencies measured in hertz denote radians per second, when the correct definition of hertz is cycles per second.

    \section{Minimal introduction to tripartite entanglement}
            \label{sec:intro-entanglement}
            Approximately two thirds of this work deals with entanglement. We ignore the percentage of scientific articles produced today that deal with quantum information, but it is certainly high enough to make it impossible for one person to keep up with all of them. Therefore, we will use the present section to make a non-exhaustive introduction to entanglement. In fact, it only covers the concepts strictly required to understand the following chapters.

            If the reader is familiar with any quantum technology, then they must have heard about so many different kinds of entanglement that it is impossible to keep a count of them: bipartite, multipartite, genuine, completely inseparable, $k$-producible, \textit{GHZ}- or \textit{W}-like, ... Behind this diversity, however, there is a fundamental reason: the definition of entanglement itself. So, lets take a step back and briefly ask \textit{What is entanglement?} The shortest definition we know of is \textit{non-separability}. In order to give meaning to the definition, consider one of the text-book postulates of Quantum Mechanics
            \begin{quote}
                \textbf{Postulate 4:} The state space of a composite physical system is the tensor product of the state spaces of the component physical systems [...]
                \begin{flushright}
                    From
                    \textit{Quantum Information and Quantum Computation}, 
                    \\
                    M. A. Nielsen and I.L. Chuang.
                \end{flushright}
            \end{quote}
            It indicates how to build new, more complicated, systems from simpler ones. That is a very convenient postulate to have. With its aid, we can split the Universe into simple constituents, figure out their dynamics when isolated from the rest, and then join some of them together and try to solve the whole thing. 

            However, when the compound system evolves in time and correlations appear among its components, a tricky situation may happen: Even if we know everything there is to know about the compound system, this is, the global state $\psi$; it might make no sense to separate pieces of the global state and consider them the states of each of the components, as it happens in $\ket{\psi} = \frac{1}{\sqrt{2}}\ket{\psi_1}\otimes\ket{\psi_2} + \frac{1}{\sqrt{2}}\ket{\phi_1}\otimes\ket{\phi_2}$. Such a phenomenon is \textit{non-separability}, and when it takes place we claim that the global state is \textit{entangled}.

            Please note the negation in its definition: There is entanglement when there is \textit{no} separable description of at least one of the constituents. And if entanglement consists in the absence of a property... How can it not be astonishingly diverse? Think, for instance, about the variety of things that are \textit{not} blue. Such a set contains elements as different as a robin and the Sahara desert, not to mention ridiculous elements to which the concept of color does not apply at all, such as the scent of wine, the dialectic process and many others. There is, however, a crucial difference between entanglement and the set of not-blue things. The definition of entanglement does have an affirmative ingredient, entanglement \textit{is} (no negation there) a property of some quantum states, which is a much tamer set to begin with that the not-blue set. Having said that, the dimension of the global state space grows exponentially with the number of subsystems, so even if it is not as incomprehensible as the \textit{not} blue set, it is still difficult to handle.
        \subsection{Multipartite entanglement}
            One of the first displays of the diversity of entangled states arises when considering tripartite systems. When the global system is made of three different constituents its state might be non-separable in more than one way
            \begin{align*}
                \psi_{123} &= 
                \frac{1}{\sqrt{2}}\psi_1\otimes\psi_2\otimes\psi_3
                + \frac{1}{\sqrt{2}}\phi_1\otimes\phi_2\otimes\phi_3
                \\
                \psi_{12-3} &= 
                \left[\frac{1}{\sqrt{2}}\psi_1\otimes\psi_2
                + \frac{1}{\sqrt{2}}\phi_1\otimes\phi_2
                \right] \otimes \psi_3
            \end{align*}
            The state $\psi_{123}$ is definitely non-separable, but the state $\psi_{12-3}$ has two separable pieces, one of which is, in turn, non-separable. Situations like this one motivated the distinction between \textit{bipartite}, \textit{tripartite} and, in general, $N$ \textit{multipartite} entanglement. In section \ref{sec:ghz-vs-w} we provide another criterion, based on whether two states can be converted one into another with only local operations and classical communication, that allows to further divide $N$ multipartite entangled states for a fixed $N$.
        \subsection{Genuine entanglement}
            \label{sec:genuine-entanglement}
            The second insight into the variety of entangled states, relevant in the chapters to come, deals with our ignorance about the actual global state. Up until this point, our definition of entanglement relied on an assumption impossible to meet: \textit{if we know everything that there is to know about the global system} is a ridiculous statement in an experimental science such as physics. In order to introduce a model of our ignorance, physicists argued that Born's rule should be modified
            \begin{quote}
                \textbf{Born's rule:} Given a system with state $\psi$ and an observable $O$, the expectation value of the latter is
                \begin{align*}
                    \expval{O} = \bra{\psi}O\ket{\psi}
                \end{align*}
            \end{quote}
            Note that given a state, the rule returns the expectation value of an observable. If instead of being certain about the state of the system, we could only be certain of sampling the state from an ensemble of orthonormal (so they are distinguishable) states, then Born's rule should be combined with classical probability theory
            \begin{quote}
                \textbf{generalized Born's rule:} Given a system with state sampled from a ensemble of orthonormal states $\{\psi_i\}_{i=1}^N$ with probabilities $P_i$, as well as an observable $O$, its expectation value is
                \begin{align}
                    \expval{O} = \sum_{i=1}^RP_i\bra{\psi_i}O\ket{\psi_i}
                    \label{eq:modified-borns-rule}
                \end{align}
            \end{quote}
            Since the sampled $\psi_i$s are orthonormal then they are a basis of the state space, or if that is not the case, the ensemble can be extended with more states with probability zero until it is a basis. In that case, the modified Born's rule becomes taking the trace of a multiplication of two operators
            \begin{align*}
                \expval{O} = Tr\{\rho O\}
            \end{align*}
            where $\rho = \sum_{i}P_i\ket{\psi_i}\bra{\psi_i}$. Any reader with basic knowledge in quantum mechanics knows this operator as the density matrix, and it is customary to interpret it as a \textit{mixed} state (even though, within this interpretation, it is not a state properly speaking) while the actual states the density matrix samples from are called \textit{pure} states. 

            Now, within this formalism, what is quantum entanglement, again? We could take a naive route and define it directly as non-separability of the density matrix. In this case, a system, even when we do not know the global state, is entangled whenever the global density matrix is not a tensor product of density matrices of each elementary subsystem, that is, $\rho = \rho_1 \otimes \rho_2$. If we were to use that definition, the following density matrix should be considered entangled and non-separable
            \begin{align*}
                \rho = 
                \frac{1}{2}\ket{00}\bra{00}
                +
                \frac{1}{2}\ket{11}\bra{11}
            \end{align*}
            But, if we take a look at the modified Born's rule in Eq. (\ref{eq:modified-borns-rule}) we can appreciate this density matrix can be interpreted as having the global state sampled from the two clearly separable possibilities $\ket{00}$ or $\ket{11}$, both with the same probability. Therefore, we should not claim this density matrix contains any entanglement, as it can be interpreted as a classically correlated distribution of separable pure states. We take this chance to define separable pure states as product states, since their density matrices are tensor products of the subsystem's matrices.

            Following with the example, we could be led to believe that the fruitful definition of separability does not pay attention to the separability of the whole matrix, but to the separability of the pure states it samples from. In that case our density matrix should be considered separable because it can be interpreted as sampling from separable pure states. However, many different probability distributions of pure states end up producing the same density matrix. In particular, the density matrix we are considering can be reinterpreted in terms of Bell pairs
            \begin{align*}
                \ket{\psi_{\pm}} &= 
                \frac{1}{\sqrt{2}}\ket{00}
                \pm
                \frac{1}{\sqrt{2}}\ket{11}
            \end{align*}
            and then it follows
            \begin{align*}
                \rho =
                \frac{1}{2}\ket{00}\bra{00}
                +
                \frac{1}{2}\ket{11}\bra{11}
                =
                \frac{1}{2}\ket{\psi_-}\bra{\psi_-}
                +
                \frac{1}{2}\ket{\psi_+}\bra{\psi_+}
            \end{align*}
            Therefore, we conclude that given a density matrix, the particular ensemble of states producing it is ambiguous and inadequate for a definition of separability. But, on the other hand, not all density matrices display such extreme ambiguity. If we know the state is a Bell pair, there is only one density matrix for the system and it is clearly entangled. Thus, somewhere between these two extreme cases must lie the border between separable and non-separable states. The actual definition of separability is as follows: a density matrix is said to be separable if there is at least one ensemble of only product states that produces it. Conversely, the definition of non-separability follows: a density matrix is said to be non-separable if there is \textit{no} ensemble of only product states that produces it. 
            \begin{align*}
                \rho_{\text{non-separable}}
                \neq
                \sum_{i=1}^{R}
                P_i
                \rho_{i}^{(1)}
                \otimes
                \rho_{i}^{(2)}
            \end{align*}

            Note that, just as it happened with pure states, the definition of entanglement is formulated as a negation, this is, as non-separability, and that opens the door to many different possible ways of being entangled. Following the steps we took with pure states, tripartite systems display some of those different kinds of entanglement. Firstly, consider a separable density matrix
            \begin{align*}
                \rho_{\text{1-2-3}} 
                =
                \sum_{i=1}^R
                P_i
                \rho_i^{(1)}
                \otimes
                \rho_i^{(2)}
                \otimes
                \rho_i^{(3)}
            \end{align*}
            Again, note that the decomposition in the right hand side of the equation is not unique, but as soon as one like this one exists, the left hand side is a separable tripartite density matrix. From here, we can try to define new non-separable density matrices. Take for example
            \begin{align*}
                \rho_{\text{12-3}}
                = \sum_{i=1}^R
                P_i
                \rho_i^{(1,2)}
                \otimes
                \rho_i^{(3)}
                \text{ with }
                \rho_i^{(1, 2)} 
                \neq
                \sum_{j=1}^{R'}
                P_{ij}
                \rho_{ij}^{(1)} \otimes \rho_{ij}^{(2)}.
            \end{align*}
            This density matrix is entangled, according to our definition, but that entanglement is shared only between the subsystems 1 and 2. Therefore, we denominate its entanglement as bipartite. Additionally, we take the opportunity offered by this example to make a second definition: biseparability. Here, the third system is separable from the composite system made from the first and second ones, and all three together form the global system. Then, we claim the state is biseparable. Such concept becomes very convenient when making the jump to tripartite entanglement. Consider a new example in which a density matrix is not biseparable in any possible way
            \begin{align}
                \rho_{\text{fully inseparable}}
                \neq \sum_{j=1}^{R'}
                P_{ij}
                \rho_{ij}^{(\alpha, \beta)}
                \otimes
                \rho_{ij}^{(\gamma)}
                \label{eq:full-inseparability}
            \end{align}
            where the superindices $\alpha$, $\beta$ and $\gamma$ indicate the subsystem those operators act upon. They range from 1 to 3 without repetition and the order between $\alpha$ and $\beta$ does not matter, so Eq. (\ref{eq:full-inseparability}) is a summarized way of imposing 3 equations on $\rho_{\text{fully inseparable}}$. Is this tripartite entanglement? The answer is that it might, but does not have to be. We call situations like these \textit{fully inseparable}, but not tripartite entanglement. To clarify this distinction consider
            \begin{align}
                \rho_{\text{generalized biseparable}}
                &=
                \frac{1}{3}
                \rho^{(1,2)}
                \otimes
                \rho^{(3)}
                + \frac{1}{3}
                \rho^{(1)}
                \otimes
                \rho^{(2,3)}
                + \frac{1}{3}
                \rho^{(1,3)}
                \otimes
                \rho^{(2)}
                \nonumber \\
                &\text{with }
                \rho^{(\alpha, \beta)}
                \neq
                \sum_{i=1}^{R}
                P_i
                \rho_i^{(\alpha)}
                \otimes
                \rho_i^{(\beta)}
                \label{eq:generalized-biseparable-example}
            \end{align}
            This density matrix is fully inseparable because there is no bipartition that separates the matrix in two. But, if we interpret the global density matrix with the modified Born's rule, we are modeling a system in which our knowledge about the state is enough to tell that there is some bipartite entanglement, but we do not have enough information to point out which pair of subsystems is entangled. Therefore, we claim the density matrix contains a kind generalized bipartite entanglement, not a tripartite one. Conversely, when the density matrix can not be brought into such a decomposition, we characterize it as \textit{genuinely} entangled in a tripartite way.
            \begin{align}
                \rho_{\text{genuine}}
                &= 
                \sum_{i=1}^{R}
                P_i\rho_{i}^{(1,2,3)}
                \text{ with }
                \rho_{i}^{(1,2,3)}
                \neq
                \sum_{\substack{j=1...R' \\ \alpha, \beta, \gamma = 1,2,3}}
                P^{\alpha, \beta, \gamma}_{ij}
                \rho_{ij}^{(\alpha)}
                \otimes
                \rho_{ij}^{(\beta, \gamma)}
                \label{eq:genuine-entanglement-definition}
            \end{align}
            For these states, and following again the interpretation provided to us by the modified Born's rule about density matrices, our knowledge of the state of the system is not absolute but it is sufficient to be certain about the tripartite nature of its entanglement, hence the name \textit{genuine tripartite entanglement}. Unfortunately, the name \textit{genuine entanglement} has been coined twice and for different concepts. Some authors use it to denote the entanglement of the \textit{GHZ} state as opposite to the entanglement of the $W$ state, a completely different notion than the one considered here. In our opinion, the name \textit{genuine} is sticking to the concept of non-generalized biseparability considered here; and to some extend, it has the right to be called genuine. Even more unfortunate is that both meanings of \textit{genuine entanglement} are relevant in the present thesis. Therefore, section \ref{sec:ghz-vs-w} deals with that second meaning, that we will denominate \textit{GHZ}-like entanglement.

            Additionally, in the literature about this topic one can find that $N>3$ multipartite genuine entanglement is defined in two different ways: Firstly, it can mean that a system with $N>3$ constituents is in a state that is not biseparable nor a convex sum of biseparable states on any possible bipartition (for instance \cite{antipin2021}). Other authors consider that an $N$-partite system can have $M$-partite genuine entanglement (with $N \geq M \geq 3$) by extending the definition recursively. That is, a system is genuinely $M$-partite entangled if its state can not be decomposed as a convex sum of $(M-1)$partite genuine entangled states (for instance \cite{guo2021genuine}). Needless to say, the first definition is the same as the second one if $M$ is 3.

            Before concluding this section, please note that tripartite genuine entanglement, despite being of fundamental interest because of its experimental robustness, is quite difficult to handle. It has a convoluted definition that indicates a system is entangled whenever, from the many different decompositions of the density matrix, there is no generalized biseparable description of it. Therefore, there are few measures of genuine entanglement, a reasonable search on arXiv (\href{https://arxiv.org/search/?query=genuine+entanglement+measure&searchtype=title&abstracts=show&order=-announced_date_first&size=50}{this one}) on January 2022 yields 15 results about genuine entanglement measures, the most relevant of them in our opinion being \cite{ma_chen2011, li_jia2017, li_wang2017, dai_dong2020, roy_das2020, guo2021b, antipin2021, guo2021genuine}. Those measures are often difficult to compute, not to mention that many do not apply to continuous variables systems as the systems we will study in the following chapters. A very common approach in the literature is, then, to compute simpler lower bounds of those measures. That way, if the bound turns out to be positive, we know the system is genuinely entangled, but not how much exactly. If the bound turns out to be negative, that is inconclusive evidence because the actual measure might be positive. An even simpler approach consists in using entanglement witnesses. Witnesses are bounds that combinations of observables of a system with no entanglement (genuine or any kind considered) must follow. If after measuring the observables, the bound is violated then the system must be entangled (in the way considered). If the bound is followed, the evidence is again inconclusive. Witnesses are, therefore, very similar to measure bounds, and they are usually easier to come up with from a theoretical perspective, and they are more accessible experimentally, as they are based on expectation values of observables. Chapters \ref{chapter1} and \ref{chapter3} will deal with the construction of these witnesses.
        \subsection{Nonequivalent tripartite entanglement classes in 3 qubit systems}
            \label{sec:ghz-vs-w}
            The third insight into the variety of phenomena we denominate entanglement deals with a broadly known result \cite{dur_vidal2000} about 3 qubit pure states.  At the beginning of section \ref{sec:intro-entanglement} we stated that entanglement is non-separability. Then, we claimed that because of its definition as the absence of a property, we are bound to have many different kinds of entanglement. That definition is perfectly fine, but it ignores a crucial fact. Quantum mechanical systems are often described effectively by Classical Mechanics if there is no specific experimental effort to \textit{produce} or \textit{protect} their quantum features. Thus, an ongoing effort in quantum research is to identify, from all the possible operations that could be performed on a system, those that generate entanglement and those that do not.

            One of the first steps in that direction could be considered the class of Local Operations and Classical Communication (LOCC) protocols, which was built to determine which operations do not generate entanglement. Its proper definition requires paying attention to some subtleties beyond the scope of this work, but in order to introduce it we indicate that, if we consider a particular separable state, we can try to find all the operations that map it to another separable state.

            The first kind of LOCC operations we consider is the unitary evolution of an isolated subsystem. If a particular subsystem is not allowed to interact with anything else, then the time evolution of the whole system factors into a tensor product containing the separable evolution of the isolated subsystem and the rest of systems. Then, we define a local unitary as the evolution of the isolated subsystem. If its initial state was separable from the rest of the system, it will remain separable. In fact, the initial state has to come from the evolution of separable states in the past. Because of this reversibility, local unitaries are known not to generate nor to destroy entanglement. If they did the latter, their inverse would do the former.

            The second kind of elementary operations that we consider in LOCC are measurements on single subsystems. These, very much like local unitaries, map the set of separable states to itself. But they are no longer reversible, and in fact, they map entangled states to separable ones. Therefore, they are said to conserve or destroy entanglement, never to create it.

            The final step to build the complete set of LOCC protocols is to consider the composition of both kinds of operations on the same subsystem or in different ones, as well as their conditional execution based on the results of previous measurements. In short, LOCC is the set of all operations that can be produced by local action on each subsystem and classical communication. There is much more to tell about LOCC protocols, but in the sake of brevity we only point out two properties. First, there are operations that map separable states to other separable states that are not in LOCC \cite{PhysRevA.59.1070}. Second, and despite the first point, LOCC protocols are often considered the complete set of non-entanglement-generating operations. Therefore, when given a general state $\rho$ we can find a LOCC protocol that generates another arbitrary state $\sigma$ from $\rho$, we say that $\sigma$ is less or equally entangled than $\rho$. If the LOCC protocol is reversible then $\sigma$ and $\rho$ are equally entangled. This argument lies at the core of the construction of every entanglement measure.

            Now, we are in the position to introduce \cite{dur_vidal2000}. It is a well known fact that Bell pairs represent the maximally entangled states of 2 qubit systems. That is, from any Bell pair we can obtain any other 2 qubit pure state by means of a LOCC operation \cite{PhysRevLett.83.436}. Furthermore, all 2 qubit states can be partially ordered from least to most entangled based on LOCC convertibility. But that is no longer the case with 3 qubit pure states. In \cite{dur_vidal2000} it was shown that the following two states
            \begin{align*}
                \ket{GHZ} &= 
                \frac{1}{\sqrt{2}}\left(
                    \ket{000} + \ket{111}
                \right) \\
                \ket{W} &=
                \frac{1}{\sqrt{3}}\left(
                    \ket{001} + \ket{010} + \ket{100}
                \right)
            \end{align*}
            can not be converted in any direction into each other by means of LOCC protocols with non-zero success probability. Even more so, any other tripartitely entangled pure state can be obtained with non-zero success probability from only one of those two states. Therefore, the authors concluded that there are two different tripartite entanglement classes, one with the $\ket{GHZ}$ state as its maximally entangled representative, and the other one lead by $\ket{W}$. As pointed out in section \ref{sec:genuine-entanglement}, it was unfortunate that some posterior literature coined the expression \textit{genuine entanglement} to the class of $\ket{GHZ}$-like states. We believe that choice was made on the grounds of how the two different states react to single system measurements. The $\ket{W}$ state becomes a Bell pair between the unmeasured qubits, while the $\ket{GHZ}$ state leaves those qubits in a separable state. Under that scope, it appears the word \textit{genuine} is applied to the $\ket{GHZ}$ entanglement because the entanglement contained in the $\ket{W}$ state is considered \textit{not genuinely tripartite}, given that it becomes bipartite under single-qubit measurement. We find this consideration misleading, the $\ket{W}$ state is tripartitely inseparable, just as the $\ket{GHZ}$ is. In fact, we find a much more appropriate term would be to denominate the entanglement of the $\ket{W}$ as \textit{robust} to particle loss, while the $\ket{GHZ}$ is \textit{fragile}. Moreover, chapter \ref{chapter3} generalizes the concept of (non-)Gaussian introduced in chapter \ref{chapter1} to discrete variables systems and, there, we will show how the $\ket{GHZ}$ must be considered non-Gaussian while the $\ket{W}$ state is Gaussian.

    \section{Resources for learning cQED}
        \label{sec:intro-cQED}
        An introduction to the broad field of circuit quantum electrodynamics (cQED) is beyond the expected contents of this thesis. Fortunately, its actual contents deal with a limited set of circuits.  But if the reader finds out they are unfamiliar with the design of superconducting circuits, we provide some guidance on interesting references for the particular set of systems that we consider.

        cQED deals with the modeling of nearly dissipativeless superconducting circuits. Quantum mechanics provides us with a microscopical explanation on why those materials contain dissipativeless currents, but their quantum properties reach much further than that. These circuits are modeled by macroscopic degrees of freedom that are quantum in nature. While a typical electronic circuit can not be in a superposition state such as \textit{an electron has gone through the device} plus \textit{that same electron has not}, a superconducting circuit can. Such quantum behaviour is due to multiple factors: from the Bose-Einstein condensation of Cooper pairs to the experimental exploitation of the Josephson effect and flux quantization, which lead to the design of Josephson junctions and superconducting quantum interference devices (SQUIDs). A very accessible introduction to these topics is provided on the last chapter of the Feynman lectures \cite{feynman-lectures}.

        Josephson junctions are of great importance to cQED because they behave as nonlinear inductors, which in turn produce anharmonic energy levels when they are included in circuits. If the reader is unfamiliar with how those circuits are modeled from its constituent Jospephson junctions, capacitors and other elements, we recommend the course on quantum fluctuations in electrical circuits, given by Devoret at Les Houches in 1995 \cite{devoret1995}. Having the ability to produce anharmonic energy levels is fundamental from a perspective of controlling the circuit's state. In anharmonic systems, electromagnetic pulses of different frequencies address transitions between different levels. This fact allows for greater control of the state than if the levels were harmonic, as it would happen with no Josephson junctions. Such control led to the design of qubits. These systems are electrical circuits that can be cooled down to its ground state and operated coherently so that its dynamics remain bounded to a state subspace of dimension two. However, there are a variety of circuits that can behave that way, leading to the multitude of different qubits we have today: charge and flux qubits, transmons, xmons and so on. A modern review such as \cite{blais_grimsmo2021} covers the circuits that are most often used nowadays, as well as many of the successes of cQED: the strong light-matter coupling of qubits to the field on waveguides, the ability of modulate over time the circuit's parameters when they include SQUIDs and many more. Ultimately, that review explains how these circuits are operated to realize a digital quantum computer. Albeit those results are of great importance, they are not relevant for this thesis. We often deal with analog simulators or with entanglement generation in systems prior to interpret them as quantum computers.

        Up to this point we have discussed cQED in a broad sense. It is a very active area of research, so keeping up with the entirety of the literature is a challenging task. Therefore, we provide now references that are tailored to this thesis, instead of trying to cover cQED completely. In the chapters to come, we often consider one-dimensional cavities terminated in a SQUID. The article \cite{eichler_wallraff2014} contains a good introduction to how to model that system, despite they only consider a single Josephson junction instead of a complete SQUID. Other systems that we consider are tunable-coupling transmons, which are introduced in \cite{srinivasan_hoffman2011}.

    \bibliographystyle{apsrev4-2}
    \bibliography{thesis.bib}{}

	\chapter{Towards the detection of genuine tripartite non-Gaussian entanglement}
	\label{chapter1}
		At the beginning of this PhD, experiments were taking place in the groups of C. M. Wilson (University of Waterloo) and G. Johansson (Chalmers University of Technology) with the theoretical collaboration of my PhD advisor, C. Sabín. They were running an experiment in which a superconducting circuit behaved as a parametric amplifier capable of down-converting a pump tone into three different tones at once \cite{chang_sabin2020}, a process denominated three-mode spontaneous parametric down-conversion (3SPDC). Loosely speaking, the vacuum state of the three output tones was evolved into a superposition of that vacuum plus the state with a photon on each mode. This just happens to be a good approximation, in general many other photon number states are involved in that evolution, but those two are the main participants at short times. Such an approximate state is reminiscent of the \textit{GHZ} state in discrete variables, and therefore we expected that the parametric oscillator, operated as this three-mode mixer, generated tripartite entanglement. Because we were modeling an experiment, we had to consider the system as being in a mixed state, and in those cases the entanglement we were expecting is denominated \textit{genuine}. Unfortunately that term has been coined twice with different meanings. Here, we use it to describe that we are certain about the tripartite nature of the entanglement even when we are not absolutely certain about the system's state, as we introduced in section \ref{sec:intro-entanglement}. But other literature has considered it as the entanglement of the \textit{GHZ} and related states, which can be confusing in this chapter, and even more so in chapter \ref{chapter3}.

        However, prior literature \cite{gonzalez_borne2018} had argued that this was not a fruitful scheme for generating genuine entanglement. In that work, a similar system in the context of nonlinear quantum optics was studied and they indicated that a pump tone that parametrically down-converted into three output tones produces a state that was not detected as genuinely entangled by a family of witnesses. They proceeded to study other pump tones and seeding schemes to successfully generate entanglement. Nevertheless, we must point out that witnesses are not entanglement measures, they are only sufficient (but not necessary) conditions to entanglement. Therefore, their argument did not rule out tripartite entanglement in the 3SPDC setting and we set out to find new witnesses families that were tailored for the experiment.

        A first analysis of the system indicated that the witnesses used in \cite{gonzalez_borne2018} were not well suited for the situation at hand because they paid attention only to covariances of the quadratures. The 3SPDC system does not change the covariances of vacuum, its action becomes apparent in higher statistical moments. Witnesses that were sensitive to those moments were not found, so we built one and it detected genuine tripartite entanglement in our system. 

        Then, we realized that all the witnesses we had gathered followed a pattern: those that paid attention only to covariances could not detect entanglement encoded in higher statistical moments and, conversely, witnesses that paid attention to those higher moments did not detect entanglement encoded in covariances. A prime example of the latter entanglement is produced in this same or related systems when the pump consists of two tones that resonate with different pairs of the three output modes. That way, the initial vacuum state in the output nodes $\ket{000}$ is driven into a superposition in which a pair of modes is excited $\ket{110}$, plus another pair excited $\ket{011}$. We denominate that process double two-mode spontaneous parametric down-conversion (2-2SPDC) and it is known to generate genuine tripartite entanglement \cite{lahteenmaki_paraoanu2016, bruschi_sabin2017, chang_simoen2018}. As pointed out above, the witness built to detect the entanglement produced by 3SPDC is oblivious to the entanglement produced in 2-2SPDC. Inspired by the field of quantum optics, we chose to coin the adjective \textit{Gaussian} for the covariance-sensitive witnesses, as those moments determine completely a Gaussian; while we coined \textit{non-Gaussian} for the witnesses composed of higher statistical moments. Since Gaussian witnesses reported entanglement only in the 2-2SPDC process, while non-Gaussian witnesses did only in the 3SPDC, we were lead to believe that these could be mutually exclusive entanglement classes, regardless of the witnesses considered, hence the expression \textit{non-Gaussian} in the title of this chapter.

        The following sections in this chapter meet three purposes: Firstly, they give an introduction to topics required to understand the results summarized above. Secondly, one of those sections is the actual publication these results lead to. And thirdly, some sections provide improvements over the results published in \cite{agusti_chang2020}. Therefore, section \ref{sec:3spdc-intro} introduces the design on the superconducting system that enables the 3SPDC process. Section \ref{sec:about-entanglement-and-witnesses} introduces our non-Gaussian witness. Section \ref{sec:3spdc-publication} is the publication \cite{agusti_chang2020}, whose introduction covers in far lesser detail the topics discussed in the previous sections, but its main body discusses the application of our non-Gaussian witness to the 3SPDC system. There, numerical simulations can be found that identify what experimentally accessible parameters regimes are adequate for genuine tripartite non-Gaussian entanglement generation. After the publication, section \ref{sec:witness-improvement} introduces an improvement of our witness that was found later on. Lastly, we provide some concluding remarks on section \ref{sec:3spdc-conclusions}.

    \section{Three-mode spontaneous parametric down conversion in circuit Quantum Electrodynamics}
        \label{sec:3spdc-intro}
        Here we describe the experimental system that we studied in \cite{agusti_chang2020} and that was introduced in \cite{chang_sabin2020}. In fact, we gather some of the result of \cite{chang_sabin2020} here. In short, what we re-derive in this section is the combination of resonators and Josephson junctions that lead to an electromagnetic field that can be effectively described by the Hamiltonian
        \begin{align*}
            H_{\text{3SPDC}} =
            \sum_{i=1}^3\omega_ia^\dagger_ia_i
            + g_0 a^\dagger_1a^\dagger_2a^\dagger_3
            + g_0a_1a_2a_3,
        \end{align*}
        that is, three normal modes with characteristic frequencies $\omega_i\quad i=1,2,3$ that interact in a three-body fashion with coupling $g_0$. The $i$-th mode creation operator is $a^\dagger_i$. Note that we set $\hbar = 1$.

        \subsection{Slightly asymmetric and weakly pumped SQUID}
            The system described in \cite{chang_sabin2020} is composed of a superconducting cavity and an asymmetric Superconducting QUantum Interference Device (SQUID) sitting at one of its edges. We start our model with the latter. As pointed out in section \ref{sec:intro-cQED}, a SQUID is a loop of superconducting material, except for two points (Josephson junctions, from now on JJ) that break the loop in two different islands that are so close together that the tunneling of Cooper pairs becomes a relevant phenomenon. Because of the Josephson effect and flux quantization through the loop, the inductance of the device depends on both the current and the external magnetic flux passing through it, enabling experimentalists to tune the device's non-linear inductance in real time. In particular, the Lagrangian is
            \begin{align*}
                L_{\text{SQUID}}(\phi_1, \dot{\phi}_1, \phi_2, \dot{\phi}_2, t) &= 
                \frac{C_1}{2}\dot{\phi}_1^2
                + E_{J1}\cos\left(
                    \frac{\phi_1}{\phi_0}
                    \right) \\
                &+ \frac{C_2}{2}\dot{\phi}_2^2
                + E_{J2}\cos\left(
                    \frac{\phi_2}{\phi_0}
                    \right),
            \end{align*}
            where $\phi_1$ is the phase difference between the Cooper pairs at each side of the first junction, whereas $\phi_2$ is the same at the second one. We denote time derivatives with the dot notation, for instance in $\dot{\phi}_i$. The capacities of each junction are $C_1$ and $C_2$, while $E_{J1}$ and $E_{J2}$ are their respective Josephson energies.  The reduced flux quantum is $\phi_0 = 1/2e$, with $1=\hbar$ the reduced Planck's constant and $e$ the (positive) electron charge. Note that the phase differences must follow the flux quantization condition
            \begin{align*}
                \phi_2 - \phi_1 = \phi_{\text{ext}},
            \end{align*}
            where $\phi_{\text{ext}}$ is the external magnetic flux through the loop and we suppose the system contains no flux quanta. Note that $\phi_{\text{ext}}$ must be considered time dependent. In fact, it is the mechanism that introduces the pump tone into the system, so it must be regarded as oscillating with the pump frequency.  For convenience, we define an independent flux $\phi = \phi_1 + \phi_{\text{ext}}/2 = \phi_2 - \phi_{\text{ext}}/2$ so that the Lagrangian might be rewritten as having only one dynamical variable. After some simple but convoluted algebra, the single-variable Lagrangian turns out to be equivalent to a single tunable Josephson junction
            \begin{align}
                L_{\text{SQUID}}(\phi, \dot{\phi}, t) &= 
                \frac{C_T}{2}\dot{\phi}^2
                +
                E_J(\phi_{\text{ext}})
                \cos\left(
                    \frac{\phi}{\phi_0} 
                    - \alpha(\phi_{\text{ext}})
                \right),
                \label{eq:exact-effective-JJ-for-asym-SQUID}
            \end{align}
            where the effective parameters $E_J$, $\alpha$ and $C_T$ can be written in terms of both JJs characteristic parameters and the external flux
            \begin{align*}
                E_J(\phi_{\text{ext}}) &=
                \sqrt{
                        E_{J1}^2 
                        + E_{J2}^2 
                        + 2E_{J1}E_{J2}\cos\left(
                            \frac{\phi_{\text{ext}}}{\phi_0}
                        \right)} \\
                \alpha(\phi_{\text{ext}}) &= 
                \text{atan}\left(
                    \tan\left(\frac{\phi_{\text{ext}}}{2\phi_0}\right)
                    \frac{E_{J1} - E_{J2}}{E_{J1} + E_{J2}}
                \right) \\
                C_T &= C_{1} + C_{2}.
            \end{align*}
            The dependence of the effective Josephson energy $E_J$ and phase offset $\alpha$ on the external flux is rather complicated, in order to simplify our analysis we consider three approximations. Firstly, the JJs in the SQUID are only slightly asymmetric
			\begin{align*}
				\Delta
				&= \frac{E_{J2} - E_{J1}}{E_{J1} + E_{J2}},
			\end{align*}
            that is, we neglect any quadratic term in $\Delta$ from the effective Josephson energy
            \begin{align}
                E_J \approx 
                2E_{J1}\sqrt{1 + 2\Delta}
                \left|\cos\left(
                    \frac{\phi_{\text{ext}}}{\phi_0}
                \right)\right|,
                \label{eq-effective-Josephson-energy1}
            \end{align}
            which now has a simpler trigonometric dependence on the external flux. The second approximation we perform involves the pump tone. The external magnetic flux oscillates around some mean value with the pump tone frequency
            \begin{align*}
                \phi_{\text{ext}}(t) = 
                \phi^0_{\text{ext}} + \lambda \cos(\omega_d t),
            \end{align*}
            where $\omega_d$ is the pump tone. We   consider that the amplitude of the oscillations is small compared to the flux quantum, that is $\lambda << \phi_0$ and we neglect any quadratic or higher term in $\lambda/\phi_0$. This way the cosine dependence of the effective Josephson energy in Eq. (\ref{eq-effective-Josephson-energy1}) can be approximated by
            \begin{align*}
                \cos\left(\frac{\phi_{\text{ext}}}{\phi_0}\right)
                &=
                \cos\left(
                    \frac{\phi^0_{\text{ext}}}{\phi_0}
                    \right)
                \cos\left(
                    \frac{\lambda\cos{\omega_dt}}{\phi_0}
                    \right) 
                - \sin\left(
                    \frac{\phi^0_{\text{ext}}}{\phi_0}
                    \right)
                \sin\left(
                    \frac{\lambda\cos{\omega_dt}}{\phi_0}
                    \right) \\
                &\approx
                \cos\left(
                    \frac{\phi^0_{\text{ext}}}{\phi_0}
                    \right)
                - \sin\left(
                    \frac{\phi^0_{\text{ext}}}{\phi_0}
                    \right)
                \frac{\lambda}{\phi_0}\cos{\omega_dt}
                + O\left(\frac{\lambda^2}{\phi_0^2}\right) \\
                &\approx
                \cos\left(
                    \frac{\phi^0_{\text{ext}}}{\phi_0}
                    \right)
                \left[
                1 - \tan\left(
                    \frac{\phi^0_{\text{ext}}}{\phi_0}
                    \right)
                \frac{\lambda}{\phi_0}\cos{\omega_dt}
                \right]
                + O\left(\frac{\lambda^2}{\phi_0^2}\right),
            \end{align*}
            which yields
            \begin{align*}
                E_J \approx 
                2E_{J1}\sqrt{1 + 2\Delta}
                \left|
                \cos\left(
                    \frac{\phi^0_{\text{ext}}}{\phi_0}
                    \right)
                \right|
                \left|
                1 - \tan\left(
                    \frac{\phi^0_{\text{ext}}}{\phi_0}
                    \right)
                \frac{\lambda}{\phi_0}\cos{\omega_dt}
                \right|,
            \end{align*}
            Additionally we define some constants in order to clarify the dependence of the effective Josephson energy on the pump tone.
            \begin{align*}
                E &= 
                2E_{J1}\sqrt{1 + 2\Delta}
                \left|
                \cos\left(
                    \frac{\phi^0_{\text{ext}}}{\phi_0}
                    \right)
                \right| \\
                \delta E &= \frac{E}{\phi_0}\tan\left(
                    \frac{\phi^0_{\text{ext}}}{\phi_0}
                    \right)
            \end{align*}
            so that the final, conveniently approximated, effective Josephson energy is
            \begin{align}
                E_J \approx 
                \left|
                E -  \lambda \delta E\cos{\omega_dt}
                \right|,
                \label{eq-effective-Josephson-energy3}
            \end{align}
            where we have kept the dependence on the pump amplitude $\lambda$ explicit, that is, outside of the definitions of $E$ and $\delta E$ because we are not done with the approximation $\lambda^2/\phi_0^2 << 1$. In particular, that approximation has further simplifying effects on the effective phase offset $\alpha(\phi_{\text{ext}})$. After some algebra and expanding a Taylor series on $\lambda$, it becomes
            \begin{align*}
                \alpha(\phi_{\text{ext}}) 
                \approx
                \text{arctan}\left(
                    \Delta
                    \tan\left(\frac{\phi_{\text{ext}}^0}{2\phi_0}\right)
                \right)
                + \frac{
                    \sec^2\phi_{\text{ext}}^0/2\phi_0\Delta
                    }{
                        1 + \tan^2\left(
                            \phi_{\text{ext}}^0/2\phi_0
                            \right)\Delta^2 
                    } \lambda \cos\left(\omega_dt\right)
                + O\left(\frac{\lambda^2}{\phi_0^2}\right),
            \end{align*}
            where the first time-independent factor can be ignored redefining the independent potential $\phi$ as the addition of the previous potential plus that constant factor. The second time-dependent factor oscillates with the pump tone and a complicated amplitude which we encapsulate in the variable $\delta\alpha$, that is equal to
            \begin{align*}
                \delta \alpha = 
                \frac{
                    \sec^2(\phi_{\text{ext}}^0/2\phi_0)\Delta
                    }{
                        1 + \tan^2\left(
                            \phi_{\text{ext}}^0/2\phi_0
                            \right)\Delta^2 
                    }.
            \end{align*}
            Note that, again, we do not include the pump amplitude $\lambda$ within the definition of $\delta\alpha$ because we are not finished taking the approximation $\lambda^2/\phi_0^2 << 1$ yet. Summarizing, the Lagrangian for the slightly asymmetric, low amplitude pump SQUID is
            \begin{align}
                L_{\text{SQUID}} \approx
                \frac{C_T}{2}\dot{\phi}^2
                + \left|E - \lambda \delta E\cos(\omega_dt)\right|
                \cos\left(
                    \frac{\phi}{\phi_0} - \lambda \delta\alpha \cos(\omega_dt)
                \right)
            \end{align}
            which has a simpler dependence on the pump tone than the full effective Lagrangian (\ref{eq:exact-effective-JJ-for-asym-SQUID}). Lastly, the third approximation that we perform consists in considering the internal flux of the SQUID, that is the dynamical variable $\phi$ itself, small in units of the quantum flux. However, we take this approximation not as strongly as the others before, and we neglect terms in the Lagrangian that behave as $O(\phi^6/\phi_0^6)$
            \begin{align*}
                L_{\text{SQUID}} &\approx
                \frac{C_T}{2}\dot{\phi}^2
                + \left|E - \lambda \delta E\cos(\omega_dt)\right| \\
                &\times\left[
                    1 -
                    \frac{1}{2}\left(
                        \frac{\phi}{\phi_0} 
                        - \lambda \delta\alpha \cos(\omega_dt)
                    \right)^2 +
                    \frac{1}{4!}\left(
                        \frac{\phi}{\phi_0} 
                        - \lambda \delta\alpha \cos(\omega_dt)
                    \right)^4
                    + O(\phi^6)
                \right]
            \end{align*}
            We ignore the zero-order term, which is constant in the dynamic variable $\phi$ and has no effect on the equations of motion. Additionally, we find the final terms that are proportional to $\lambda^2$ and regard them as very small
            \begin{align}
                L_{\text{SQUID}} \approx
                \frac{C_T}{2}\dot{\phi}^2
                &+\lambda E\delta\alpha \cos(\omega_dt)\frac{\phi}{\phi_0}
                -\frac{
                    \left|E - \lambda \delta E \cos(\omega_dt)\right|
                  }{2}
                 \frac{\phi^2}{\phi^2_0}
                \nonumber \\
                &-\frac{\lambda E\delta\alpha}{6} \cos(\omega_dt)
                 \frac{\phi^3}{\phi^3_0}
                +\frac{
                    \left|E - \lambda \delta E \cos(\omega_dt)\right|
                 }{24}\frac{\phi^4}{\phi^4_0}
                \label{eq:squid-lagrangian}
            \end{align}
            Note this Lagrangian can be split in a linear and a non-linear part in $\phi$ the dynamic variable
            \begin{align}
                \label{eq:squid-linear-lagrangian}
                L_{\text{SQUID linear}} =
                \frac{C_T}{2}\dot{\phi}^2
                &+\lambda E\delta\alpha \cos(\omega_dt)\frac{\phi}{\phi_0}
                -\frac{
                    \left|E - \lambda \delta E \cos(\omega_dt)\right|
                  }{2}
                 \frac{\phi^2}{\phi^2_0}
                \\
                L_{\text{SQUID non-linear}} =
                &-\frac{\lambda E\delta\alpha}{6} \cos(\omega_dt)
                 \frac{\phi^3}{\phi^3_0}
                +\frac{
                    \left|E - \lambda \delta E \cos(\omega_dt)\right|
                 }{24}\frac{\phi^4}{\phi^4_0}
                \label{eq:squid-non-linear-lagrangian}
            \end{align}
            Summarizing, we have proven that a slightly asymmetric and weakly pumped SQUID can be modeled as a time-dependent non-linear inductor joined by a conventional capacitor. In the following section we will describe the effects that such an inductor can have on a cavity when it is built at one of its edges.

        \subsection{Effects of the slightly asymmetric and weakly pumped SQUID on a one-dimensional cavity}
            Now, we turn our attention to the complete system, which is composed of said SQUID at the edge of a one-dimensional cavity. This cavity is described by a magnetic flux field $\phi(x, t)$, with the coordinate $x$ spanning from $x = 0$ to $x = d$, being $d$ the cavity length. What we named $\phi$ as the SQUID's internal magnetic flux is now the field at the edge, $\phi(x = d, t)$. However, before writing down the system's Lagrangian, we must point out that the transition from a discrete description of the SQUID to a continuous field when considering the cavity must be paid some attention. In particular, we will begin with a discrete description of the cavity in terms of $N$ coupled and equally spaced \textit{LC}-resonators
            \begin{align*}
                L_{\text{resonators}} =
                \sum_{i=1}^N\frac{c\Delta x\dot{\phi}^2_i}{2}
                + \sum_{i=1}^{N-1}\frac{(\phi_{i+1}-\phi_{i})^2}{2l\Delta x},
            \end{align*}
            where $\Delta x = \frac{d}{N-1}$ is the spacing between the resonators and $c$ and $l$ are the capacitance and inductance of the cavity per unit of length. The magnetic flux at the $i$-th resonator is $\phi_i$ and it can be thought of as an approximation to the magnetic flux field at $\phi(x = i \Delta x, t)$. In fact, it converges to it when the jump to a field description is given, that is, $N \rightarrow \infty$ and $\Delta x \rightarrow 0$ while keeping $(N-1)\Delta x = d$ constant.

\begin{figure}
			\centering
			\includegraphics[scale=1]{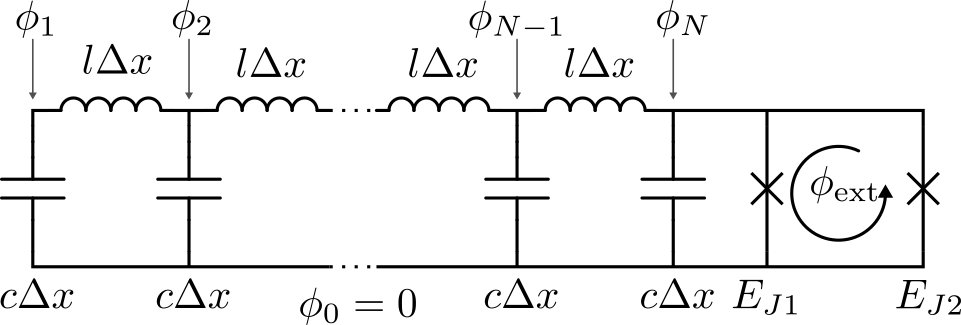}
			\caption{Circuit representing the discrete lumped element approximation model for a one-dimensional cavity ending on a slightly asymmetric weakly-pumped SQUID. The cavity is split into $N$ identical LC resonators with flux variables $\phi_i$, each of them accounting for a fractional piece $l \Delta x$ of the total inductance $l d$, and another fractional piece $c \Delta x$ of the total capacitance $c d$. Two Josephson junctions with characteristic energies $E_{J1}$ and $E_{J2}$ form the SQUID which introduces a parametric dependance on the external magnetic flux $\phi_{\text{ext}}$ going through its loop.}
			\label{fig:circuit}
			\end{figure}

            In Figure (\ref{fig:circuit}) we show a schematic of the lumped element circuit that approximates the cavity field and how we connect the SQUID to the system. In short, we connect it to the last node, $\phi_N$, which will become the field at the edge of the cavity $\phi(d, t)$ when we take the continuum limit. Then, the complete Lagrangian is
            \begin{align}
                L_{\text{total}} &= 
                L_{\text{resonators}} 
                + L_{\text{SQUID linear}} + L_{\text{SQUID non-linear}}
                \nonumber \\
                &= \sum_{i=1}^N\frac{c\Delta x\dot{\phi}^2_i}{2}
                + \sum_{i=1}^{N-1}
                \frac{(\phi_{i+1}-\phi_{i})^2}{2l\Delta x} 
                \nonumber \\
                &+\lambda E\delta\alpha \cos(\omega_dt)\frac{\phi}{\phi_0}
                -\frac{
                    \left|E - \lambda \delta E \cos(\omega_dt)\right|
                  }{2}
                 \frac{\phi^2}{\phi^2_0}
                \nonumber \\
                &-\frac{\lambda E\delta\alpha}{6} \cos(\omega_dt)
                 \frac{\phi^3}{\phi^3_0}
                +\frac{
                    \left|E - \lambda \delta E \cos(\omega_dt)\right|
                 }{24}\frac{\phi^4}{\phi^4_0}
                 \label{eq:total-undiagonalized-lagrangian}
            \end{align}
            where we have performed a fourth approximation: considering that the plasma frequency of the SQUID is much larger that any of the characteristic frequencies of system. That way, the SQUID self-capacitance $\frac{C_T}{2}\dot{\phi}_N^2$ can be safely ignored.

            Even after that approximation, the Lagrangian produces complicated equations of motion, in particular non-linear time-dependent ones. We could, in principle, ignore those equations, perform a Legendre transformation to get the Hamiltonian of the system and quantize its canonical variables in order to get a quantum description of the system. Then, we could try to solve a complicated quantum system to find out whether three-mode spontaneous parametric down-conversion takes place in it. However, there is a simpler procedure: we can solve the linear time-independent piece of the equations of motion and then quantize the system. This way, the eigen-modes obtained in this classical description become bosonic modes and we can focus our attention only on the non-linear and/or time-dependent terms at the quantum level. Therefore, it is worthwhile to take a look at the equations of motion having ignored the non-linear terms and taken the time average of the time-dependent ones
            \begin{align*}
                \ddot{\phi}_1 &= 
                -\frac{\phi_1 - \phi_2}{cl \Delta x^2} \\
                \ddot{\phi}_i &= 
                -\frac{2\phi_i-\phi_{i-1}-\phi_{i+1}}{cl \Delta x^2} \\
                \ddot{\phi}_N &=
                -\frac{\phi_{N-1} - \phi_N}{cl \Delta x^2}
                + E\frac{1}{2}\frac{\phi_N}{\phi_0^2}
            \end{align*}
            when the resonators' fluxes are interpreted as values of the field, those equations become
            \begin{align}
                \ddot{\phi}(0, t) &= 
                -\frac{1}{cl\Delta x}\left[
                \frac{\phi(0, t) - \phi(\Delta x, t)}{\Delta x}
                \right]
                \label{eq:first-boundary-origin} \\
                \ddot{\phi}(x, t) &= 
                -\frac{1}{cl}\left[
                \frac{
                    2\phi(x, t)
                    -\phi(x -\Delta x, t)
                    -\phi(x +\Delta x, t)}
                    {\Delta x^2} 
                \right]
                \label{eq:wave-eq-origin} \\
                \ddot{\phi}(d, t) &=
                \frac{1}{c\Delta x}\left[
                -\frac{\phi(d - \Delta x, t) - \phi(d, t)}
                {l \Delta x}
                +\frac{E}{\phi_0 }
                \frac{\phi(d, t)}{2\phi_0}
                \right]
                \label{eq:last-boundary-origin}
            \end{align}
            where $x$ is not a continuous variable yet, but rather $x = i\Delta x$. However, when the field limit is taken, Eq. (\ref{eq:wave-eq-origin}) becomes the well known wave equation. That same limit exists for Eq. (\ref{eq:first-boundary-origin}) only if the divergence of the factor $\frac{1}{cl\Delta x}$ is compensated by a null second factor, which gives the open boundary condition $\partial_x \phi(0, t) = \phi'(0, t) = 0$. Once that condition is satisfied, the equation becomes again the wave equation. Performing the same analysis on Eq. (\ref{eq:last-boundary-origin}), the boundary condition at $x = d$ turns out to be
            \begin{align*}
                \frac{\phi'(d, t)}{l} 
                -\frac{E_J}{\phi_0}
                \frac{\phi(d, t)}{2\phi_0} = 0
            \end{align*}
            Then, we can propose eigen-modes that implicitly follow the boundary condition at $x = 0$, that is $\phi_n(x) = A_n \cos(k_nx)$, which in turn transform the boundary condition at $x = d$ into a transcendental equation yielding the wave numbers
            \begin{align*}
                k_nd\tan(k_nd) = \frac{ldE}{2\phi_0^2}.
            \end{align*}
            The best we can do is numerically finding out the values of $k_n$ given the systems parameters. Considering those wave numbers as known, the general solution to the wave equation $\phi(x, t) = \sum_{n}\phi_n(t)\cos(k_nx)$ allows us by direct integration to write down the linear time-average Lagrangian as
            \begin{align*}
                L_{\text{linear time-averaged}} = 
                \sum_n\left[
                    c_n\dot{\phi}_n^2(t) + l_n^{-1}\phi^2_n(t)
                \right]
            \end{align*}
            with an effective capacitance and inductance per $n$-th mode that can be expressed as
            \begin{align*}
                c_n &= 
                c\int_0^ddx\cos^2\left(k_nx\right)
                = \frac{cd}{2}
                  \left(1 + \frac{\sin\left(2k_nd\right)}{2k_nd}\right) \\
                l_n^{-1} &= l^{-1}k_n^2\int_0^ddx\sin^2\left(k_nx\right)
                + \frac{E_J}{\phi_0^2}\cos\left(k_nd\right)
                = \frac{k^2_nd^2}{2ld}\left(
                1 - \frac{\sin\left(2k_nd\right)}{2k_nd}
                \right)
            \end{align*}
            Now, it is time to pay attention to the time-dependent and/or non-linear terms that we ignored in the total Lagrangian in Eq. (\ref{eq:total-undiagonalized-lagrangian}). There is a linear time-dependent term that becomes:
            \begin{align*}
                E\delta\alpha\cos\left(\omega_dt\right)\frac{\phi(d, t)}{\phi_0} = \lambda\cos(\omega_d t)\sum_nM_n^{(1)}\phi_n(t),
            \end{align*}
            where we have grouped the constant coefficients into $M_n^{(1)}$ for convenience, its definition is
            \begin{align}
                M_n^{(1)} =
                \frac{E\delta\alpha}{\phi_0}\cos\left(k_nd\right)
                \label{eq:m1}
            \end{align}
            If we interpret each mode $\phi_n$ as a harmonic oscillator, this Lagrangian term represents a classical drive, that is, a force, acting on each mode individually. In the next section we will see that the pumping tone $\omega_d$ will not resonate with any of the modes, and therefore this driving will be negligible. Back in Eq. (\ref{eq:total-undiagonalized-lagrangian}) we see that we ignored another Lagrangian term that produced linear time-dependent terms in the equations of motion
            \begin{align*}
                \frac{
                    \lambda \delta E \cos(\omega_d t)     
                }{2}\frac{\phi^2}{\phi^2_0}
                =
                \lambda\cos(\omega_dt)
                \sum_{n, m} M^{(2)}_{n, m}\phi_n(t)\phi_m(t)
            \end{align*}
            which can be interpreted as a pair-wise interaction between the modes, driven with the same pump tone when the indices $n$ and $m$ are different. When they are equal, that term can be interpreted as a time-dependent Kerr non linearity. Again, we have grouped together the constant coefficients into a new symbol $M_{n, m}^{(2)}$ for convenience
            \begin{align}
                M^{(2)}_{n, m} = \frac{E}{2\phi_0^2}\cos(k_nd)\cos(k_md)
                \label{eq:m2}
            \end{align}
            Now it is time to turn our attention to the non-linear terms in the total Lagrangian. We split them in three different terms: one time-dependent and proportional to $\phi^3(d, t)$, another time-independent and proportional to $\phi^4(d, t)$ and a final one time-dependent and proportional to $\phi^4$ as well (see Eq. (\ref{eq:total-undiagonalized-lagrangian})). Those terms become now
            \begin{align*}
                \frac{\lambda E \delta\alpha}{6} \cos(\omega_dt)
                \frac{\phi^3}{\phi^3_0}
                &= \lambda\cos(\omega_dt)
                \sum_{n, m, l} M^{(3)}_{n, m, o} 
                \phi_n(t)\phi_m(t)\phi_o(t)
                \\
                \frac{E}{24}\frac{\phi^4}{\phi^4_0}
                &= \sum_{n, m, o, p}N^{(4)}_{n, m, o, p}
                \phi_n(t) \phi_m(t) \phi_o(t) \phi_p(t)
                \\
                \frac{
                    \lambda \delta E \cos(\omega_dt)
                }{24}\frac{\phi^4}{\phi^4_0}
                &= \lambda \cos(\omega_d t)
                \sum_{n, m, o, p}M^{(4)}_{n, m, o, p}
                \phi_n(t) \phi_m(t) \phi_o(t) \phi_p(t)
            \end{align*}
            where, again, we have grouped together the constant coefficients in the new symbols $M^{(3)}_{n, m, o}$ and $M^{(4)}_{n, m, o, p}$, which are defined as
            \begin{align}
                M^{(3)}_{n, m, o} &= 
                \frac{E\delta\alpha}{6\phi_0}
                \cos(k_n d) \cos(k_m d) \cos(k_o d)
                \label{eq:m3}
                \\
                N^{(4)}_{n, m, o, p} &= 
                \frac{E}{24\phi_0^4}
                \cos(k_n d) \cos(k_m d) \cos(k_o d) \cos(k_p d)
                \\
                M^{(4)}_{n, m, o, p} &= 
                \frac{\delta E}{24\phi_0^4}
                \cos(k_n d) \cos(k_m d) \cos(k_o d) \cos(k_p d)
                \label{eq:m4}
            \end{align}
            Summing up we can write the total Lagrangian in terms of the classical modes as
            \begin{align}
                L_{\text{total}} 
                &=
                \sum_n\left(
                    \frac{c_n}{2}\dot{\phi}^2_n(t) 
                    - \frac{1}{2l_n}\phi_n^2(t)
                    - \lambda \cos(\omega_d t) M^{(1)}_n\phi_n(t)
                \right)
                \nonumber \\
                &-\lambda\cos(\omega_d t)\sum_{n, m} 
                M^{(2)}_{n, m}\phi_n(t)\phi_m(t)
                \nonumber \\
                &+\lambda\cos(\omega_d t)\sum_{n, m, o}
                M^{(3)}_{n, m, o}\phi_n(t)\phi_m(t)\phi_o(t)
                \nonumber \\
                &+\sum_{n, m, o, p} N^{(4)}_{n, m, o, p}
                \phi_n(t)\phi_m(t)\phi_o(t)\phi_p(t)
                \nonumber \\
                &-\lambda\cos(\omega_dt)
                \sum_{n, m, o, p} M^{(4)}_{n, m, o, p}
                \phi_n(t)\phi_m(t)\phi_o(t)\phi_p(t)
                \label{eq:total-diagonal-lagrangian}
            \end{align}
            Summarizing, a slightly asymmetric and weakly pumped SQUID built at the edge of an otherwise open one-dimensional cavity introduces a tunable classical driving, as well as tunable two-, three- and four-mode interactions. Additionally it introduces quadratic, cubic and quartic non-linearities on single modes as well as pairs and triplets of them. Although this complete Lagrangian seems very complicated, it is ready to be quantized and then, after an astute choice of the pump tone, it will greatly simplify into a three-mode spontaneous parametric down-conversion Hamiltonian.

        \subsection{Quantization and the Rotating Wave Approximation}
            Following the standard procedure, the conjugate momenta $\varphi_n(t)$ to the magnetic fluxes $\phi_n(t)$ are introduced by means of $\varphi_n(t) = \partial L_{\text{total}} / \partial \dot{\phi}_n = c_n\phi_n$. Then, the Legendre transform of the Lagrangian returns the Hamiltonian $H = \sum_n\varphi_n\dot{\phi}_n - L_{\text{total}}$ resulting in
            \begin{align}
                H_{\text{total}} 
                &=
                \sum_n\left(
                    \frac{\varphi_n^2(t)}{2c_n}
                    + \frac{\phi_n^2(t)}{2l_n}
                    + \lambda \cos(\omega_d t) M^{(1)}_n\phi_n(t)
                \right)
                \nonumber \\
                &+\lambda\cos(\omega_d t)\sum_{n, m} 
                M^{(2)}_{n, m}\phi_n(t)\phi_m(t)
                \nonumber \\
                &-\lambda\cos(\omega_d t)\sum_{n, m, o}
                M^{(3)}_{n, m, o}\phi_n(t)\phi_m(t)\phi_o(t)
                \nonumber \\
                &-\sum_{n, m, o, p} N^{(4)}_{n, m, o, p}
                \phi_n(t)\phi_m(t)\phi_o(t)\phi_p(t)
                \nonumber \\
                &+\lambda\cos(\omega_dt)
                \sum_{n, m, o, p} M^{(4)}_{n, m, o, p}
                \phi_n(t)\phi_m(t)\phi_o(t)\phi_p(t)
                \label{eq:total-classical-hamiltonian}
            \end{align}
            Now, the conjugate variables $\phi_n$ and $\varphi_n$ can be quantized by turning them into operators on a Hilbert space fulfilling the canonical commutation relations $[\varphi_n, \phi_m] = i\hbar\delta_{n, m}$. In that quantum description we can introduce creation and annihilation bosonic operators
            \begin{align}
                \phi_n =
                \sqrt{\frac{1}{2}\sqrt{\frac{l_n}{c_n}}}
                \left(a^\dagger_n + a_n\right)
                \label{eq:annihilation-operator-definition}
            \end{align}
            so that $[a_n, a^\dagger_m] = \delta_{n, m}$ and $[a_n, a_m] = [a^\dagger_n, a^\dagger_m] = 0$. In terms of those ladder operators the Hamiltonian is
            \begin{align}
                H_{\text{total}} 
                &=
                \sum_n\left(
                    \omega_na^\dagger_na_n
                    + \lambda \cos(\omega_d t) \tilde{M}^{(1)}_n
                      (a^\dagger_n + a_n)
                \right)
                \nonumber \\
                &+\lambda\cos(\omega_d t)\sum_{n, m} 
                \tilde{M}^{(2)}_{n, m}
                (a^\dagger_n + a_n)
                (a^\dagger_m + a_m)
                \nonumber \\
                &-\lambda\cos(\omega_d t)\sum_{n, m, o}
                \tilde{M}^{(3)}_{n, m, o}
                (a^\dagger_n + a_n)
                (a^\dagger_m + a_m)
                (a^\dagger_o + a_o)
                \nonumber \\
                &-\sum_{n, m, o, p} 
                \tilde{N}^{(4)}_{n, m, o, p}
                (a^\dagger_n + a_n)
                (a^\dagger_m + a_m)
                (a^\dagger_o + a_o)
                (a^\dagger_p + a_p)
                \nonumber \\
                &+\lambda\cos(\omega_dt)
                \sum_{n, m, o, p}
                \tilde{M}^{(4)}_{n, m, o, p}
                (a^\dagger_n + a_n)
                (a^\dagger_m + a_m)
                (a^\dagger_o + a_o)
                (a^\dagger_p + a_p)
                \label{eq:total-quantum-hamiltonian}
            \end{align}
            where the tildes on the symbols $\tilde{M}^{(1)}_n$, $\tilde{M}^{(2)}_{n, m}$, ... mean that they include the constant coefficients that relate the creation-annihilation operators to the canonical variables in Eq. (\ref{eq:annihilation-operator-definition}).

            In this quantum description, the equation of motion is the Schrödinger equation. In principle, finding its solution from Hamiltonian (\ref{eq:total-quantum-hamiltonian}) is a challenging process, given the multipartite and time-dependent nature of the interactions. However, an astute choice of the pumping tone can greatly simplify our analysis. It is a well-known fact in the phenomenon of parametric down-conversion that the pump tone must be equal to the sum of the output tones, so that energy is conserved. Therefore, if we are looking after three-mode parametric down-conversion processes, the pump tone $\omega_d$ must be equal to the sum of three of the mode frequencies $\omega_n$. For simplicity and convenience we can choose those modes to be the three in lowest energy, that is, $\omega_d = \sum_{i=1}^3\omega_n$. Under that assumption, the rotating wave approximation (RWA) indicates that most of the terms in Eq. (\ref{eq:total-quantum-hamiltonian}) produce little to no dynamics, at least at short times. We remind the reader that the RWA indicates that the Hamiltonian terms that dominate the system's dynamics are those that are constant in time in the interaction picture. To make the transition to that picture clearer, let us introduce a compact notation to indicate products of creation and annihilation operators
            \begin{align*}
                a^s_n = 
                \begin{cases}
                a^\dagger_n \quad &\text{if } s = +1 \\
                a_n \quad &\text{if } s = -1
                \end{cases}
            \end{align*}
            This way, the Hamiltonian can be conveniently rewritten in the interaction picture. Note that the effect this change of picture has on the Hamiltonian is adding a oscillating phase to each $a^s_n$ operator.
            \begin{align}
                H^{\text{int}}_{\text{total}}(t)
                &=
                \lambda \cos(\omega_d t)
                \sum_{\substack{n \\ p}}
                \tilde{M}^{(1)}_n
                e^{i(\pm\omega_d + p\omega_n)t}
                a^p_n
                \nonumber \\
                &+\lambda
                \sum_{\substack{n, m\\ p, q}} 
                \tilde{M}^{(2)}_{n, m}
                e^{i(\pm\omega_d + p\omega_n + q\omega_m)t}
                a^p_na^q_m
                \nonumber \\
                &-\lambda
                \sum_{\substack{n, m, l \\ p, q, r}}
                \tilde{M}^{(3)}_{n, m, l}
                e^{i(\pm\omega_d + p\omega_n + q\omega_m + r\omega_l)t}
                a^p_na^q_ma^r_l
                \nonumber \\
                &-\sum_{\substack{n, m, l, k \\ p, q, r, s}}
                \tilde{N}^{(4)}_{n, m, l, k}
                a^p_na^q_ma^r_la^s_k
                e^{i(p\omega_n + q\omega_m + r\omega_l + s\omega_k)t}
                \nonumber \\
                &+\lambda
                \sum_{\substack{n, m, l, k \\ p, q, r, s}}
                \tilde{M}^{(4)}_{n, m, l, k}
                e^{i(\pm\omega_d 
                    + p\omega_n + q\omega_m + r\omega_l + s\omega_k)t}
                a^p_na^q_ma^r_la^s_k
            \end{align}
            Thus, finding the constant terms is the same as finding the roots of the exponents, that is, which combination of indices $p, q, r, s$ makes each exponent zero, provided that the normal frequencies $\omega_n$ and the pump tone $\omega_d$ are known. Given the anharmonicity of the normal frequencies, we can assume that none of them are a multiple of each other. Therefore, the exponents that we find to have roots are those composed of subtracting pairs of equal frequencies, that is, degenerate solutions such as $\omega_1 - \omega_1$ and the like. Notice that we have chosen the driving to be $\omega_d = \omega_1 + \omega_2 + \omega_3$. Because of this pump tone, none of the terms containing $M^{(1)}$, $M^{(2)}$ or $M^{(4)}$ are constant: they always have an anharmonic non-zero frequency/ies left out in the exponent. But some terms containing $M^{(3)}$ and $N^{(4)}$ have zero exponent, which we consider to be the ones that dominate the dynamics and compose the RWA Hamiltonian
            \begin{align}
                H^{\text{int}}_{\text{RWA}} 
                &=
                -\frac{6\lambda}{2}
                \tilde{M}^{(3)}_{1, 2, 3}
                \left(a^\dagger_1 a^\dagger_2 a^\dagger_3
                + a_1 a_2 a_3\right)
                \nonumber \\
                &-4\sum_{\substack{n, m \\ p, q}} 
                \tilde{N}^{(4)}_{n, n, m, m}
                a^p_n
                (a^p_n)^\dagger
                a^q_n
                (a^q_n)^\dagger
                \label{eq:rwa+kerr-nonnormal-hamiltonian}
            \end{align}
            where we indicate with $H^{\text{int}}_{\text{RWA}}$ that we are dealing with the Hamiltonian from Eq. (\ref{eq:total-quantum-hamiltonian}) in the interaction picture and after performing the RWA. The first term in Eq. (\ref{eq:rwa+kerr-nonnormal-hamiltonian}) is the Hamiltonian that we are looking for, one that produces three-mode spontaneous parametric down-conversion, as can be seen from the fact that it maps the vacuum state $\ket{000}$ to the three-photon state $\ket{111}$ on each mode. However, there are some fourth-order non-linearities, the terms containing $N^{(4)}$, that are relevant according to the RWA. Luckily, they add up to a constant, as the reader can easily prove by rewriting that summation in normal ordering. \footnote{that is, creation operators to the left, followed by annihilation operators to the right} Then, the system is described by
            \begin{align}
                H^{\text{int}}_{\text{RWA}} 
                &=
                -\frac{6\lambda}{2}
                \tilde{M}^{(3)}_{1, 2, 3}
                \left(a^\dagger_1 a^\dagger_2 a^\dagger_3
                + a_1 a_2 a_3\right)
                \label{eq:rwa-hamiltonian}
            \end{align}

            Summarizing, we have proven that a system composed of a one-dimensional cavity with an open edge and a slightly asymmetric, weakly pumped SQUID at the other edge is effectively described by a three-mode spontaneous parametric down-conversion Hamiltonian if the pump is conveniently tuned to match the addition of the three lowest normal modes of the cavity. Note that the three-body interaction in Eq. (\ref{eq:rwa-hamiltonian}) could, in principle, generate an entangled state from vacuum.
        
    \section{Entanglement witness construction}
        \label{sec:about-entanglement-and-witnesses}
        In the previous section we studied a system that is governed by a three-body interaction we denominated three-mode spontaneous parametric down-conversion (3SPDC). In this section we give a brief motivation about why such interaction should produce genuine tripartite entanglement and then proceed to build entanglement witnesses that are capable of reporting the presence of that kind of entanglement. It is advisable that the reader not familiar with genuine entanglement reads our introduction to the topic on section \ref{sec:genuine-entanglement}.
        \subsection{Genuine entanglement witness}
            \label{sec:witness-construction}
            In section \ref{sec:genuine-entanglement} we defined what tripartite genuine entanglement is: the property of mixed states not decomposable into biseparable mixtures. Here, we will develop a witness tailored to detect the genuine entanglement produced by the 3SPDC interaction described in section \ref{sec:3spdc-intro}. As pointed out at the beginning of this chapter \ref{chapter1}, a witness-based methodology was already used for an analogous system in quantum optics \cite{gonzalez_borne2018} and led to inconclusive results. Therefore, we had to improve that approach, and a first analysis of the system concluded that the non-Gaussian nature of the 3SPDC interaction must be taken into account. We remind the reader that in the field of quantum optics, and all the quantum technologies that have followed, a state is called \textit{Gaussian} if its Wigner function is a Gaussian. In other words, a continuous variable state is consider Gaussian if the first and second statistical moments resulting from measuring its canonical variables are enough to determine it. In order to illustrate why the 3SPDC interaction makes non-Gaussian states, consider a first-order perturbative expansion on the Schrödinger equation with the Hamiltonian in Eq. (\ref{eq:rwa-hamiltonian}) and taking the initial state as the vacuum
            \begin{align}
                \psi(t) \approx \ket{000} + g_0t\ket{111}
                \label{eq:first-order-3SPDC-state}
            \end{align}
            at short times $t$ and low couplings $g_0$, where $\ket{n}$ is the static Hamiltonian eigenstate populated with $n$ photons and $\ket{nnn}$ is $\ket{n}\otimes\ket{n}\otimes\ket{n}$. On one hand, this approximate state is genuinely entangled. To prove this, note that it is a pure state, so its associated density matrix has only one possible decomposition. Furthermore, that density matrix is clearly non-biseparable on every bipartition. On the other hand, a simple calculation shows that the covariances of the modes' quadratures for the approximate state in Eq. (\ref{eq:first-order-3SPDC-state}) are zero
            \begin{align*}
                \Delta^2 x_ix_j &=
                \expval{x_ix_j} - \expval{x_i}\expval{x_j} = 0
                \\
                \Delta^2 p_ip_j &=
                \expval{p_ip_j} - \expval{p_i}\expval{p_j} = 0
            \end{align*}
            with $i \neq j$, so these are not variances but covariances. We consider this fact compelling evidence of the non-Gaussianity of the 3SPDC interaction. Now, consider the family of witnesses used in \cite{gonzalez_borne2018}, which was introduced in \cite{van-loock_furusawa2003} and takes the form
            \begin{align*}
                S = 
                +\min{
                    \substack{i, j, k = 1, 2, 3 \\ i \neq j \neq k \neq i}
                }\left(
                    |h_ig_i| + |h_jg_j + h_kg_k|
                \right)
                -\sum_{i, j = 1}^3g_ig_j\Delta^2x_ix_j
                - \sum_{i, j = 1}^3h_ih_j\Delta^2p_ip_j
            \end{align*}
            where the six $g_i$ and $h_i$ are free real parameters. Whenever $S > 0$, then the state is genuinely entangled. But if the covariances are zero, then $S$ will never be positive, regardless of the values taken by the free parameters. We conclude that this witness, which we denominate Gaussian because it is only sensitive to covariances, is not fit to detect the entanglement of the system if there is any.

            Then, we set out to construct non-Gaussian witnesses. A promising candidate was introduced in \cite{hillery_dung2010}. That work proposed a family of witnesses tailored to detect $N$ multipartite inseparability. In particular, given two operators $O_1$ and $O_2$ acting on two different subsystems, if the condition
            \begin{align*}
                |\expval{O_1O_2}|
                > 
                \sqrt{\expval{O_1^\dagger O_1}\expval{O_2^\dagger O_2}}
            \end{align*}
            is fulfilled, then the (probably mixed) state comprising the systems labeled $1$ and $2$ is not separable. This condition is well suited for our system because it allows us to pick the operators so that the expectation values are third order (and above) statistical moments of the canonical variables. That way, the condition becomes sensitive to the state's non-Gaussianity. Consider that we set $O_1 = a_1$ and $O_2 = a_2a_3$, where $a_i$ is the destruction operator on the $i$-th mode. In this case, if we define
            \begin{align}
                I_1 = 
                |\expval{a_1a_2a_3}|
                -
                \sqrt{
                    \expval{a_1^\dagger a_1}
                    \expval{a_2^\dagger a_2 a_3^\dagger a_3}
                    }
                \label{eq:inseparability1-23-witness}
            \end{align}
            then whenever $I_1 > 0$ we know the first mode is not separable from the composite system comprising the second and third modes. If we go back to the approximated state in Eq. (\ref{eq:first-order-3SPDC-state}) and compute $I_1$  we get $|g_0t| - g_0^2t^2$, which is bigger than zero at short times and low couplings, precisely the perturbative regime in which Eq. (\ref{eq:first-order-3SPDC-state}) works. Note that other combinations of destruction operators can be used to build $I_2$ and $I_3$, witnesses that prove the inseparability of the second and third modes from the rest, respectively.

            Those witnesses and their perturbative values look promising, but they prove only the inseparability of each of the modes from the other two. In order to build a genuine entanglement witness we need to mirror the derivation of $I_1$ in Eq. (\ref{eq:inseparability1-23-witness}) or any typical witness. Those derivations often start assuming the state does not posses the kind of entanglement we are looking for. Then, some bounds are derived that those unentangled states must necessarily follow. Because of elementary logic, if we take any other state and it turns out to violate the bounds, then it must be entangled. In other words, the violation of the bound is a sufficient condition to entanglement. Thus, what do non-genuine entangled states look like? If we look back to section \ref{sec:genuine-entanglement}, those states can range from completely separable to generalized biseparable states like in Eq. (\ref{eq:generalized-biseparable-example}). All those possibilities can be summarized as the density matrices that allow at least one decomposition of the form
            \begin{align}
                \rho
                &=
                P^{1-23}
                \sum_{i = 1}^{R}
                P_i^{1-23}
                \rho^{(1)}_i
                \otimes
                \rho^{(2, 3)}_i
                \nonumber \\
                &+
                P^{2-13}
                \sum_{i = 1}^{R'}
                P_i^{2-13}
                \rho^{(2)}_i
                \otimes
                \rho^{(1, 3)}_i
                \nonumber \\
                &+
                P^{3-12}
                \sum_{i = 1}^{R''}
                P_i^{3-12}
                \rho^{(3)}_i
                \otimes
                \rho^{(1, 2)}_i
                \label{eq:non-genuine-entangled-state}
            \end{align}
            where the density matrices $\rho^{(\alpha, \beta)}_i$ might or might not be biseparable in return.

            If we consider a state like Eq. (\ref{eq:non-genuine-entangled-state}) then the expectation value of $a_1a_2a_3$ in absolute value is bounded from above by the expectation value of each of the three possible biseparable pieces of $\rho$, which we name
            \begin{align*}
                \rho^{1-23} &= 
                \sum_{i = 1}^{R}
                P_i^{1-23}
                \rho^{(1)}_i
                \otimes
                \rho^{(2, 3)}_i
                \\
                \rho^{2-13} &= 
                \sum_{i = 1}^{R'}
                P_i^{2-13}
                \rho^{(2)}_i
                \otimes
                \rho^{(1, 3)}_i
                \\
                \rho^{3-12} &= 
                \sum_{i = 1}^{R''}
                P_i^{3-12}
                \rho^{(3)}_i
                \otimes
                \rho^{(1, 2)}_i
            \end{align*}
            so that the total density matrix is their convex sum, and the expectation value of $a_1a_2a_3$ follows
            \begin{align}
                &|\expval{a_1a_2a_3}_\rho| 
                \nonumber \\
                &\leq
                P^{1-23}|\expval{a_1a_2a_3}_{\rho^{1-23}}|
                +
                P^{2-13}|\expval{a_1a_2a_3}_{\rho^{2-13}}|
                +
                P^{3-12}|\expval{a_1a_2a_3}_{\rho^{3-12}}|
                \label{eq:inefficient-witness-bound}
            \end{align}
            because of the triangle inequality. After the publication of \cite{agusti_chang2020} we realized there is a tighter and simpler bound, which we introduce in section \ref{sec:witness-improvement}. In order to make a genuine non-Gaussian witness from this bound we need to take a couple more steps. Firstly, the density matrices $\rho^{\alpha-\beta\gamma}$ are at least biseparable by construction. Therefore, we know each of them make $I_\alpha \leq 0$, so that
            \begin{align*}
                |\expval{a_1a_2a_3}_{\rho^{\alpha-\beta\gamma}}|
                \leq
                \sqrt{
                    \expval{
                        a_\alpha a_\alpha^\dagger
                    }_{\rho^{\alpha-\beta\gamma}}
                    \expval{
                        a_\beta a_\beta^\dagger
                        a_\gamma a_\gamma^\dagger
                    }_{\rho^{\alpha-\beta\gamma}}
                }
            \end{align*}
            Secondly, we need to drop any dependence on the biseparable density matrices. Those are not the actual state of the system and checking any possible decomposition of a state looking for them is unpractical. Luckily, for any operator we have
            \begin{align*}
                P^{1-23}\expval{O}_{\rho^{1-23}}
                =
                \expval{O}_\rho
                -
                P^{2-13}\expval{O}_{\rho^{2-13}}
                -
                P^{3-12}\expval{O}_{\rho^{3-12}}
                \leq
                \expval{O}_\rho,
            \end{align*}
            and similarly for the other permutations of the subsystems. This inequality points out that the expectation value over one of the biseparable matrices composing the total mixed state will never exceed the expectation value of the actual mixed state. That way we can drop any reference to the biseparable density matrices in the bound to $|\expval{a_1a_2a_3}|$. Summarizing, any state ranging from completely separable to generalized biseparable, or in other words, not genuinely entangled, necessarily follows
            \begin{align*}
                |\expval{a_1a_2a_3}|
                \leq
                \sqrt{
                    \expval{
                        a_1 a_1^\dagger
                    }
                    \expval{
                        a_2 a_2^\dagger
                        a_3 a_3^\dagger
                    }
                }
                +
                \sqrt{
                    \expval{
                        a_2 a_2^\dagger
                    }
                    \expval{
                        a_1 a_1^\dagger
                        a_3 a_3^\dagger
                    }
                }
                +
                \sqrt{
                    \expval{
                        a_3 a_3^\dagger
                    }
                    \expval{
                        a_1 a_1^\dagger
                        a_2 a_2^\dagger
                    }
                }.
            \end{align*}
            Therefore, we conclude that violating this bound is a sufficient condition to genuine entanglement. Grouping its terms together, we define our non-Gaussian genuine witness
            \begin{align}
                &G_{\text{1}}
                = 
                |\expval{a_1a_2a_3}|
                \nonumber \\
                &-
                \sqrt{
                    \expval{
                        a_1 a_1^\dagger
                    }
                    \expval{
                        a_2 a_2^\dagger
                        a_3 a_3^\dagger
                    }
                }
                -
                \sqrt{
                    \expval{
                        a_2 a_2^\dagger
                    }
                    \expval{
                        a_1 a_1^\dagger
                        a_3 a_3^\dagger
                    }
                }
                -
                \sqrt{
                    \expval{
                        a_3 a_3^\dagger
                    }
                    \expval{
                        a_1 a_1^\dagger
                        a_2 a_2^\dagger
                    }
                }.
                \label{eq:G}
            \end{align}
            So that when $G_{\text{1}} > 0$, the state has to be genuinely entangled. Note that because of the triplet and quartets of annihilation operators, the witness is indeed sensitive to third and fourth order statistical moments of the canonical variables, that is, it is non-Gaussian.

    \section{Publication}
        \label{sec:3spdc-publication}
        In this section the article \cite{agusti_chang2020} is copied verbatim, as the main section of the present chapter. Please note it has its own page numbering as well as bibliography.
        \cleardoublepage
        \includepdf[pages=-]{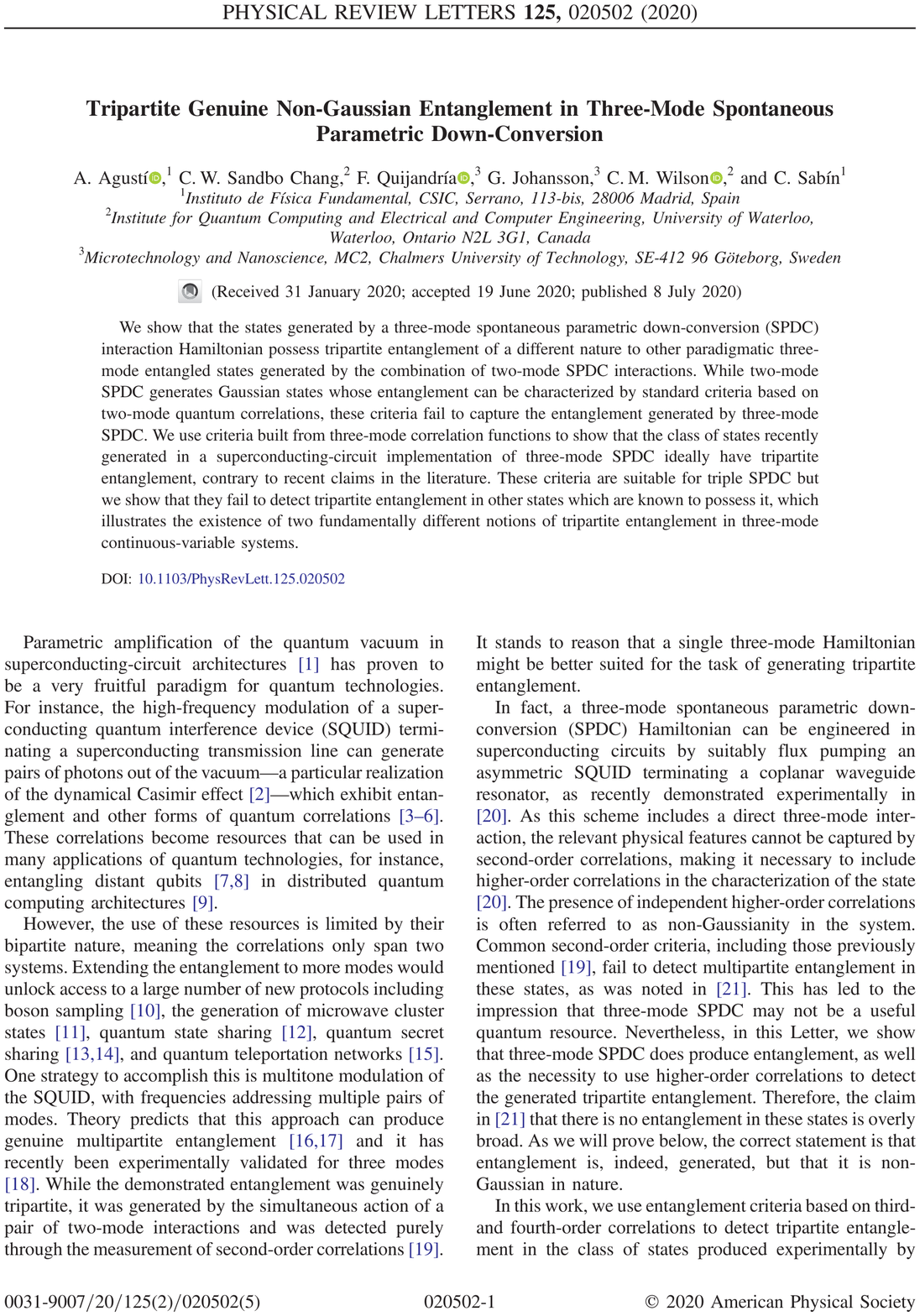}
        \cleardoublepage
    \section{Non-Gaussian witness improvement}
        \label{sec:witness-improvement}
        The reader at this point might be wondering why the sections before the article \cite{agusti_chang2020} derived the non-Gaussian witness $G$ just to have that same derivation repeated in the article itself. The reason is that the redundancy simplifies the discussion of an improvement we found on that witness, as we have pointed out a couple times previously. The key insight that made possible the improvement is a rather trivial fact: a convex sum follows the triangle inequality, yes, but it also follows the generally tighter bound that it can not be bigger than the largest of its individual terms. In other words,
        \begin{align*}
            \Big|\sum_{i=1}^R P_i v_i\Big| \leq \sum_{i=1}^RP_i|v_i|
        \end{align*}
        is true, but a better bound is
        \begin{align*}
            \Big|\sum_{i=1}^R P_i v_i\Big| \leq \max_{i=1...R}|v_i|
        \end{align*}
        Therefore, this can be applied to the construction $G_{\text{1}}$ back in section \ref{sec:witness-construction}. There, we studied a chain of bounds for non genuine states of the shape in Eq. (\ref{eq:non-genuine-entangled-state}). Those states are in general convex sums of biseparable density matrices, and we used the triangle inequality to bound the expectation value of the operator $a_1a_2a_3$ in Eq. (\ref{eq:inefficient-witness-bound}). If we use the better bound for convex sums, we have
            \begin{align}
                |\expval{a_1a_2a_3}_\rho| 
                \leq
                \max_{\substack{
                    \alpha, \beta, \gamma = 1, 2, 3 \\
                    \alpha \neq \beta \neq \gamma \neq \alpha
                }}
                |\expval{a_1a_2a_3}_{\rho^{\alpha-\beta\gamma}}|
                \label{eq:efficient-witness-bound}
            \end{align}
            and then the same steps as in section \ref{sec:witness-construction} build the following witness
            \begin{align}
                G_{\text{2}}
                =
                |\expval{a_1a_2a_3}|
                -
                \max_{\substack{
                    \alpha, \beta, \gamma = 1, 2, 3 \\
                    \alpha \neq \beta \neq \gamma \neq \alpha
                }}
                \sqrt{\expval{N_\alpha}\expval{N_\beta N_\gamma}}
                \label{eq:G2}
            \end{align}
            We claim this witness is an improvement because it is clear from the argument above that there might be states that are reported as genuinely entangled by $G_{\text{2}}$ but not from $G_{\text{1}}$.
    \section{Conclusions}
        \label{sec:3spdc-conclusions}
        In this chapter we have described our research on the feasibility of the detection of genuine entanglement in a 3SPDC process. In particular, we have given a detailed derivation of the results published in \cite{agusti_chang2020}, in the hope of increasing its accessibility to a broader audience than the article itself. We conclude that a non-Gaussian witness is capable of detecting genuine entanglement in experimentally accessible parameter regimes, and we raise to the readers attention the fact that Gaussian witnesses appear to be better suited for the Gaussian states produced by 2-2SPDC processes, while the non-Gaussian witness is better suited for the non-Gaussian states produced by the 3SPDC process.

        The reader might be wondering how far that mutual exclusion between Gaussian and non-Gaussian witnesses can be taken from a theoretical perspective. Do Gaussian and non-Gaussian genuinely entangled states define some notion of different entanglement classes? During the development of the research showcased in the Thesis we became interested in that question, which lead to a different publication that is the center of chapter \ref{chapter3}. Thus, we refer the reader to that chapter for further information.
    \bibliographystyle{apsrev4-2}
    \bibliography{thesis.bib}{}

	\chapter{Qubit Motion as a Microscopic Model for the Dynamical Casimir Effect}
	\label{chapter2}
    One of the heralded applications of quantum technologies is a radical improvement on the simulation of quantum systems. As the pioneers of quantum computing noted \cite{manin1980, manin_translated2007, feynman1982}, the very same fact that makes quantum simulation difficult, that is, the exponential growth of memory and operations required to store and compute the time evolution of a quantum system of increasing size; is the reason why we should consider simulators governed by quantum mechanics. As more physical elements are added to the simulator, its accessible resources grow exponentially too, so they compensate the growth in the computing costs of a bigger system to simulate. Once this idea was formalized \cite{lloyd1996}, the proposal, design and demonstration of quantum simulators became a very active research area. However, as quantum technologies grew more capable, multiple paradigms for performing quantum simulations have risen.

    Universal fault-tolerant quantum computers, provided one is built, are capable of efficient quantum simulation. When those computers are designed within the circuit formalism we say they execute \textit{digital} simulations. We have engaged in the design of some of those digital simulations, see \cite{cordero2021} for further details. Needless to say, devices designed today are Noisy Intermediate Scale Quantum (NISQ) computers \cite{preskill2018}, so digital simulations are far from being executed on fault-tolerant machines. It remains an open question whether there is an error mitigation scheme that will allow a NISQ device to outperform classical supercomputers on simulation problems, while there are claims of having obtained that quantum advantage on sampling tasks \cite{arute_arya2019, zhong_wang2020}.

    Alternatively, and inspired by the history of classical computing, devices that are not Turing complete, but rather mimic the physical processes of the systems to be simulated, are denominated \textit{analog} simulators. In these devices, there is a map (from now on, the analogy) between the magnitudes of interest in the simulated system and the actual experimental magnitudes of the simulator. Then, the device is designed so that the simulator's magnitudes have, in principle, the same dynamics as the magnitudes of the simulated system. This strategy places less requirements on the device than universal fault-tolerant computation, which makes it an attractive research area today. Making any claims on quantum advantage delves into details of complexity theory that are beyond the scope of this thesis, but it is undeniable that many analog simulators have produced insightful results in many-body systems \cite{bernien_schwartz2017} or even high energy physics \cite{tan_becker2021}, to name a few.

    Additionally, we believe it is worthwhile to point out two other benefits of analog simulators beyond quantum exponential growth of available resources.
    When a system is expected to exhibit some interesting phenomenon, but it is not experimentally accessible, a common strategy is to consider analog simulators that reproduce that same phenomenon on a different setting. 
    Therefore, fundamental knowledge is gained both about the phenomenon, despite the inaccessibility of the simulated system, and about the physics of the simulator, which now displays the phenomenon of interest. This idea is further developed in more concrete examples in section \ref{sec:analog-simulators}.

    On this chapter we deal with the design of an analog simulator for relativistic and quantum effects. The system that we explore contributes to present day quantum technologies in both ways:
    On one hand, it exploits the accessibility of the strong coupling regime in circuit QED to simulate relativistic effects. Note that these effects, in their original formulation in free space, often suffer from weak coupling and small signals. 
    In particular, we propose a simulator for the internal degrees of freedom of a mirror undergoing relativistic motion, which in turn is expected to produce photons, a phenomenon denominated dynamical Casimir effect. Surprisingly, we find that a simulator expected to reproduce another relativistic phenomenon, the Unruh effect, is capable of such task. Both effects are briefly introduced in section \ref{sec:relativistic-effects}, and the publication containing the results in section \ref{sec:publication-chapter2}.
    On the other hand, this simulator makes use of experimental proposals for qubits with time-dependent couplings.
    Some of these qubits were built with the intention of reducing crosstalk in digital quantum computers, so finding applications for them in analog simulators increases their interest. Other of these systems have been proposed for introducing time-dependent couplings driven by mechanical means. In the Appendix to the publication in section \ref{sec:publication-chapter2} we summarize the benefits and weaknesses of each proposal. Then, in section \ref{sec:chapter2-conclusion}, we present the conclusions and possible future directions of research.
    \section{Analog simulators}
        \label{sec:analog-simulators}
        A characteristic that sets analog simulators apart from digital computers, and very relevant to understand this chapter, is the varying transparency the analogy can have between the simulator and the simulated system. Sometimes, the system to be simulated is expected to exhibit some phenomenon and, when the simulator reproduces the same phenomenon with high fidelity, we claim the phenomenon has happened on the simulator instead.
        
        To speak in less abstract terms, allow us to consider an uncontroversial example that any undergraduate student has heard of: It is a well known fact that a spring, manipulated in the parameter regime where Hooke's law stands, is a mechanical harmonic oscillator. Likewise, an LC resonator operated within the lumped-element approximation is an electrical harmonic oscillator. Therefore, there is an analogy between the length of the spring and the voltage across the circuit, as well as relations between their parameters. This example is enlightening when we consider the discovery of the phenomenon of resonance. It is beyond the point of this thesis to make this example historically accurate, but it stands to reason that, at some point in History, someone was the first person to both observe and to have a theoretical model of the phenomenon of resonance. Let us speculate that the system considered was mechanical. Then, the system was driven with a force oscillating at the same frequency as the oscillations, which in turn rapidly grew in amplitude. Later on, the theory of electrical circuits is developed and LC resonators are designed. When driven with an oscillating power source the system undergoes an analogue growth in amplitude. The point of this example is: can we claim that we are dealing with two distinct phenomena, mechanical and electrical resonance, or we must claim they are the same phenomenon?
        
        There might be no definitive answer to that question. An interesting perspective on it pays attention to how we use the word \textit{simulator}. The difference between a simulator and its simulated system often implies that the former is more accessible than the later. At the beginning, mechanical systems were more common, so they could be considered simulators for the more rare electrical circuits. Nowadays, on the other hand, assembling LC resonators from premanufactured elements is simpler for students than dealing with the fragility of mechanical oscillators. So it is understandable to build, for example, coupled LC resonators to simulate toy models of solids, instead of the mechanical experiment. Given the fact that the preferred system is a matter of convenience beyond any physical criterion, we are inclined to claim that resonance is a phenomenon common to both systems. But, what if one of the pair of analogous systems is much more difficult to build?
        
        This is the case for the Dynamical Casimir effect (DCE). It will be briefly introduced in section \ref{sec:dce}, but for now consider that its original formulation described the generation of photons when a mirror is undergoing relativistic velocities and accelerations. Needless to say, actually building that system is a difficult technological feat, that has not been achieved at time of writing. However, one can understand that the mirror is relevant to the electromagnetic field because it provides a time-dependent boundary condition. There are many other different systems that can undergo time-dependent boundary conditions. In particular, C. M. Wilson \textit{et al} \cite{wilson_johansson2011} built, within the context of circuit QED, a one-dimensional cavity with a SQUID sitting at one of its edges so that it acts as a time-dependent inductor. In that system, the relativistic movement of the mirror is replaced by the fast tuning of the inductor, creating an analog simulator of the DCE. Photon generation was found and the DCE was claimed to be observed. Or should we say \textit{electrical} DCE? We believe we find ourselves in the same situation as in the example above and, given that some research groups are after optomechanical experiments producing the DCE \cite{macri_ridolfo2018}, then the main distinction between the electrical or mechanical DCE could be which equipment a particular laboratory has, provided the mechanical DCE is observed at some point. Therefore, the reader must be aware that when we discuss analog simulations, we make no distinction between the simulated and simulator's effects. In fact, depending on context, we refer to the effect as some sort of equivalence class between all analogous systems, or to the particular experiment under consideration.
    
        Summarizing, the design of analog simulators widens our understanding of the simulated phenomena and increases the applications current technologies have. This benefit adds up to the exponential growth in resources that set the field of quantum simulation in motion, and is the one this chapter pays most attention to.
	\section{Relativistic effects in Quantum Field Theories}
        \label{sec:relativistic-effects}
        Up to this point we have discussed quantum analog simulations from a perspective centered on the simulator. It is understandable to have such a bias nowadays, as building those simulators is an exciting new area of research, to which the article at section \ref{sec:publication-chapter2} belongs. In this section we will balance that bias and briefly explore the perspective taken by the theoretical physicists that explore relativistic effects on quantum field theories for their own sake.

        Quantum field theories are one of the most successful formalisms we have so far. They are the primary ingredient of many, if not all, theories that need to take into consideration both quantum and relativistic features, and succeed in making many experimentally contrasted predictions. Take, for instance, the magnetic moment of the electron, which is claimed to be the most accurately known property of an elementary particle \cite{odom_hanneke2006}. That magnitude is in agreement with its theoretical prediction all the way to the thirteenth decimal place. Thus, because of this experiment and many others, we are very confident about the soundness of standard model of particle physics.

        There are, however, predictions that have not been measured yet, mostly due to the difficulty of designing experiments providing a signal strong enough to be measured, or observations sensitive enough to detect them. The most famous example could be considered the Hawking radiation, with its associated difficulty of taking place at the event horizon of a black hole. Given that requirement, it is accessible only to observations and not to experiments, not to mention the difficulty of those observations. Furthermore, such phenomenon bears special interest because takes current quantum field theories to parameter regimes near to where they are expected to fail. That is, in intense gravitational fields where general relativity must be taken into account. We conclude that somewhere in between this extreme example and the tamer experiments cited above, new physics could be found. 

        Because of the equivalence principle in general relativity, we know that gravity is locally equivalent to acceleration. During the 1970s, Fulling \cite{fulling1973}, Davies \cite{davies1975} and Unruh \cite{unruh1976} considered an accelerating observer running quantum experiments in otherwise flat and empty space-time. They concluded that a detector accelerating with the observer would detect photons, even though the state of the field, as seen from a resting detector, was vacuum. This phenomenon was coined the Unruh effect, and it is further discussed in \ref{sec:unruh-effect}. Alternatively, there are other mechanisms that can introduce acceleration into quantum systems. That is the case of an accelerating mirror, which was studied by G. Moore \cite{moore1970} and later coined as the Dynamical Casimir effect (DCE). It is further discussed in \ref{sec:dce}.

        \subsection{Unruh effect}
            \label{sec:unruh-effect}
            As explained above, Davies \cite{davies1975} and Unruh \cite{unruh1976} noted that a detector experiencing constant proper acceleration through vacuum should obtain a reading equivalent to being at rest in the presence of a thermal field. Prior to that, Fulling \cite{fulling1973} had noted that the situation was more intricate than what the previous statement leads us to believe. When one considers the quantization of a field beyond Minkowski space-time, be it on actually curved space-time or on an accelerated coordinate system, there is some ambiguity when it comes to define particle number operators. Therefore, the authors cited so far concluded that what should be interpreted as the number operator for an inertial observer is not the same that for an accelerating one. Then, for the same field state the inertial and accelerated observers can detect different number of particles.

            That ambiguity in the particle number led to some authors to believe that the reading on the detector was an artifact, instead of representing the state of the field \cite{ford_oconnell2006}. Others claim that the field is, indeed, populated by photons following a thermal distribution but not on the thermal state, rather on a state produced from spontaneous emission by the excited detector \cite{scully_kocharovsky2003}. We conclude that our understanding of the effect is incomplete, so building an analog simulator could be a good way of gaining insight into these questions. We try to keep a broad enough definition of the Unruh effect until said unknowns are solved. For instance, we do not require the field to be in a thermal state. Firstly, because the founding articles on the topic only show that the response of the accelerated detector is as if the field was thermal when they are at rest, and the authors do not prove that the field is actually on \textit{the} thermal state. Secondly, even in those pioneering formalisms different detector trajectories are bound to produce a different system evolution in which the reading of the detector and the actual state of the field become something completely different from thermal.

            Following a similar line of thought, Scully \textit{et al.} \cite{scully_kocharovsky2003} considered enhancing the signal of the Unruh effect by accelerating two-level detectors into and through a high \textit{Q} cavity. In that setting, the signal is several orders of magnitude larger and the field state evolves further away from vacuum, which led them to believe this could be a step towards the detection of the effect. A key insight of \cite{scully_kocharovsky2003} is that, both in free space and in the cavity, detector and field become excited because of the dynamics produced by counter rotating terms in the Hamiltonian. We remind the reader that the rotating terms are defined as those that are constant in the interaction picture for time independent Hamiltonians, and they are counter rotating otherwise. An example will be useful throughout this chapter. Consider the Rabi Hamiltonian
            \begin{align*}
                H_{\text{Rabi}} = 
                \omega a^\dagger a 
                + \frac{\Omega}{2} \sigma_z
                + g \sigma_x (a^\dagger + a)
            \end{align*}
            where $\omega$ is the characteristic frequency of an oscillator with creation-annihilation operators $a$ and $a^\dagger$, respectively. The qubit's characteristic frequency is $\Omega$, while $\sigma_x$ and $\sigma_z$ are its first and third Pauli matrices. The coupling between the qubit and the oscillator is $g$. When this Hamiltonian is written in the interaction picture, its dependency on time can be explicitly expressed if one considers the qubit's ladder operators $(\sigma^+ + \sigma^-) = \sigma_x$
            \begin{align*}
                H_{\text{Rabi}}^{\text{int}}
                = g \left[
                    \sigma^- a^\dagger
                    + \sigma^+ a^\dagger e^{-i(\omega + \Omega)t}
                    + \sigma^- a e^{-i(\omega - \Omega)t}
                    + \sigma^+ a
                \right]
            \end{align*}
            We bring to the reader's attention that the term $\sigma^+a^\dagger$ is counter rotating, and it contributes to the excitation of both the two-level detector and the field. When the qubit is accelerating through space-time, we must consider the coupling $g$ as dependent on the qubit's position and its proper time, in order to account for the different eigen-mode amplitudes it will find throughout its trajectory. Then, the insight of \cite{scully_kocharovsky2003} consists in studying how the time-modulation of the coupling can break the Rotating Wave Approximation (RWA) so that the dynamics produced by $\sigma^+a^\dagger$ become dominant. Scully \textit{et al.} claim that this hint applies beyond the cavity-enhanced effect.

            We have concluded the introduction about the Unruh effect. The introduction is minimal, in the sense that it covers the works required to understand the article at \ref{sec:publication-chapter2}, and it overlooks the rest of the literature. This is intentional, since the scope of this thesis deals with analog simulators, and our discussion of the subtleties of the effect must end at some point. Please allow us to abandon the perspective of a physicist working on relativistic phenomena, and let us take the one centered on analogue simulators.

            As stated above, the Rabi Hamiltonian models the interaction of a two-level atom interacting with a single field mode. When the atom moves, the coupling must be regarded as time-dependent. Therefore, if a superconducting circuit is capable of changing the qubit-field coupling without actually moving the qubit, we can interpret it as a simulator. This approach was taken in \cite{felicetti_sabin2015} by Felicetti \textit{et al.}, including my PhD advisor, and it studied a system similar to the one we will discuss in \ref{sec:publication-chapter2}
        \subsection{Dynamical Casimir effect}
            \label{sec:dce}
            The Dynamical Casimir effect (DCE) is closely related to the Unruh effect. When one studies the quantization of a field under the presence of a moving mirror, a similar ambiguity in the particle number operator appears as the one observed by Fulling in the Unruh effect. Such a task was pioneered by Moore \cite{moore1970}, and later on developed by Fulling and Davies \cite{fulling_davies1976}. Moore concluded that, after solving said ambiguity, if the initial field state was the vacuum, the movement of the mirror would change the notion of particle number operators to a point were the system would appear as populated by photons. Needless to say, this is a relativistic phenomenon, so non-relativistic mirror velocities produce an amount of photons of no experimental interest. Accelerating a massive mirror to relativistic trajectories was considered a difficult experimental feat, but some proposals exist \cite{macri_ridolfo2018}.

            However, if one looks at the works cited so far, the mirror is modeled by a boundary condition. In fact, Fulling and Davies \cite{fulling_davies1976} propose the boundary condition to be an event horizon of a black hole in some circumstances. Therefore, from the perspective of quantum technologies, we could consider many different mechanisms to impose time-dependent boundary conditions in the context of quantum technologies. This approach was explored in \cite{johansson_johansson2009, johansson_johansson2010, wilson_johansson2011}, leading to a very similar system to the one studied in chapter \ref{chapter1}: a SQUID sitting at the edge of a cavity imposes a time-dependent boundary condition and becomes an electrical analogue of the DCE. Another successful approach considered using a metamaterial with tunable refractive index so that the optical length of the cavity is modulated over time \cite{lahteenmaki_paraoanu2013}.

            We have seen that if the DCE is considered as the result of time-dependent boundary conditions, and not just relativistic massive mirrors, it becomes a versatile phenomenon that explains the behavior of different parametric systems. But from the relativistic point of view, its original formulation presents a number of inconveniences. Firstly, there are no rigid solids in relativistic theories. However, the mirror appears to be one. Therefore, the original formulation of the DCE should be considered a limiting and unattainable case \cite{bosco_lindkvist2019}. In other words, it could be of fundamental interest to consider mirrors \textit{made of some matter}, so that the absence of rigidity can be studied. Secondly, the matter composing the mirror does not get entangled with the field. If it did, the time-evolution of the DCE could be non-unitary. Thirdly, if we are to measure the mechanical DCE at some point, it is beneficial to know which matter does not take energy from the field. If it did, DCE photon detection would be more difficult.

            All the concerns exposed in the previous paragraph could be addressed with the creation of a \textit{microscopic} model of the DCE. We use the word \textit{microscopic} after the so-called microscopic models of Ewald and Oseen \cite{ewald_hollingsworth1970, oseen1915} for refraction and transmission in classical optics. In those models, the constitutive relations for dielectrics are derived from the Maxwell equations in vacuum and considering the dielectric material as a lattice of electric dipoles. In other words, it was shown that a plane wave incident on the lattice will make the dipoles absorb and emit radiation so that they produce two plane waves that propagate in the bulk of the lattice. One of those waves interferes destructively with the incident wave, while the other is interpreted as the transmitted wave, which travels though the bulk at a lower speed giving birth to the refractive index of a material from microscopic considerations. This suggestive process led to de Melo e Souza \textit{et al.} \cite{souza_impens2018} to propose a microscopic model for the DCE. In their article, a single atom moving through open space replaces the time-dependent boundary condition as the representative of the mirror. Despite their system not answering all the concerns expressed above, we consider it is a very relevant step forward. With the work presented at section \ref{sec:publication-chapter2} we aim at creating a microscopic model that sheds light on some of those concerns in the context of the DCE within a cavity, as well as proposing analog electrical and mechanical simulators for it.
	\section{Publication}
        \label{sec:publication-chapter2}
        In this section the article \cite{agusti_garcia2021} is copied verbatim, as the main section of the present chapter. Please note it has its own page numbering as well as bibliography.
        \includepdf[pages=-]{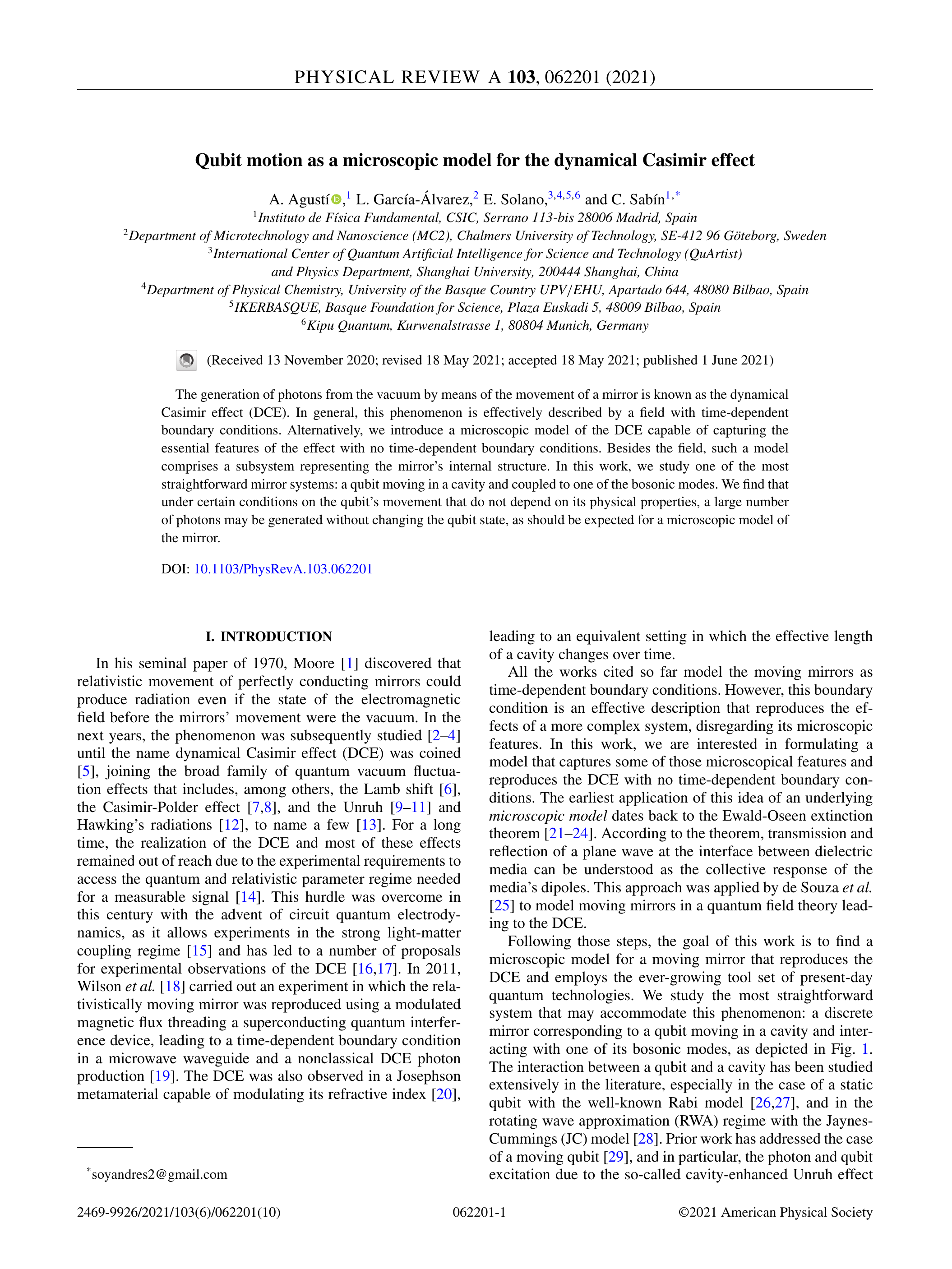}
	\section{Conclusion}
        \label{sec:chapter2-conclusion}
        In this section we briefly add to the conclusions already presented in the article at \ref{sec:publication-chapter2}. Namely, we believe it is worthwhile to interpret some of the results obtained outside of the analog interpretation.

        We have proven that a two-level system moving through a cavity at the speed of light in the medium or around that velocity, produces a monotonically increasing amount of pairs of photons while staying at its ground state and not getting entangled with the field. That is a very satisfactory microscopic model of the DCE as it follows the requisites indicated at the beginning of the article and at the end of section \ref{sec:dce}. Then we proposed a number of mechanisms that could be analog simulators for the effect. Here, we wish to add that, as with any analog simulator, it is worthwhile to interpret the phenomena at both sides of the analogy.

        From a purely electrical point of view, and abandoning any interpretation of the system as a moving mirror, the results obtained can be summarized as follows: A qubit with the ability of tuning its coupling to a waveguide is capable of being operated on four interesting regimes. Three of them were already known in previous literature. Firstly, the qubit can keep a constant strong coupling with the field, allowing for typical quantum information processing. Secondly, the qubit can decouple from the field, so that its state is protected from leaking to the waveguide. Switching between these two regimes was the original motivation behind their design \cite{srinivasan_hoffman2011}. Thirdly, the qubit can modulate its coupling so that the counter-rotating dynamics become relevant, which can be considered as a primitive to generate qubit-field entanglement. Lastly, in this work we present that a qubit can modulate its coupling to make both rotating and counter rotating dynamics relevant, leading to a large photon production while keeping the qubit at its initial state.
    \bibliographystyle{apsrev4-2}
    \bibliography{thesis.bib}{}

	\chapter{Non-Gaussian entanglement swapping between continuous and discrete variables systems}
	\label{chapter3}
    In chapter \ref{chapter1} we developed an entanglement witness tailored to the states produced by vacuum amplification from three-mode spontaneous parametric down-conversion (3SPDC). There, we found a suggestive criterion to classify entanglement in continuous variables (CV) systems: if the covariances change in time and the state remains Gaussian, then a Gaussian witness will detect the entanglement if there is any \cite{adesso2007}. If the state stops being Gaussian, considering non-Gaussian witnesses is often fruitful. Moreover, those two scenarios seem mutually exclusive. But, why should we stop at CV systems? The notions of (non-)Gaussianity discussed in the introduction of chapter \ref{chapter1} are perfectly suitable for discrete variables (DV) systems. In fact, they can be applied to hybrid systems too. Furthermore, since these entanglement notions can be applied to mixed states, we find them very flexible, as they make sense on most if not all quantum systems.
    
    Because of their general applicability, in this chapter \ref{chapter3} we pay theoretical attention to the differences and similarities of entangled states detected with (non-)Gaussian witnesses. Here, we upgrade the notions of (non-)Gaussianity in order to talk, not about witnesses nor states, but about entanglement itself. We define \textit{Gaussian entanglement} as the property contained in systems that can be detected as entangled with Gaussian witnesses. Conversely, we define \textit{non-Gaussian entanglement} as the property of systems which require to be detected with non-Gaussian witnesses. Then, we ask the question of whether there are states with both Gaussian and non-Gaussian entanglement. We conclude that 3SPDC processes produce only non-Gaussian entanglement, strengthening our claims from chapter \ref{chapter1}.

    In order to take advantage of the generality of the (non-)Gaussian criterion we study DV systems too. When considering multipartite DV systems there is a broadly known result that comes to mind. There are two classes of 3 qubit pure states that are both tripartitely entangled, and yet they are not convertible one in the other by means of stochastic local operations and classical communication (SLOCC) \cite{dur_vidal2000}, see introduction on the topic at section \ref{sec:ghz-vs-w} for further details and definitions. We find that the maximally entangled states of those classes, namely the $GHZ$ and $W$ states, contain non-Gaussian and Gaussian entanglement, respectively. This parallelism between (non-)Gaussianity and SLOCC (in-)convertibility suggest that the latter, more mathematically-oriented, classification can be interpreted in terms of the former, more experimentally-accessible, criterion.

    To put that parallelism further to the test, we study a hybrid system, that is, a system composed of both CV and DV elements. Our interest on hybrid systems is motivated by the flexibility of (non-)Gaussian entanglement and, additionally, the technological significance of these systems. We remind the reader that most qubits are in fact CV systems designed so that a two-dimensional state subspace contains their dynamics. Then, necessarily, the degrees of freedom interacting with the qubit are still CV. These CV systems can be considered the qubits' environment, or they can be engineered to couple qubits together. Thus, we find interest in modifying the original 3SPDC system. When the 3SPDC process is driven between the lowest frequency modes of three different resonators, instead of modes on the same resonator, it produces the same non-Gaussian entanglement as in chapter \ref{chapter1} and, additionally, qubits can be coupled to each resonator. The absorption of the 3SPDC radiation by the qubits leads to tripartite entangled states in those qubits that we prove to be non-Gaussian in nature. Therefore, the non-Gaussian entanglement is swapped from CV to DV systems. Moreover, the qubits are reported as entangled in a wider parameter regime. Both of these facts could enhance the technological relevance of non-Gaussian entanglement.

    Summarizing, the present chapter revolves around the results published in \cite{agusti_sabin2022}, which is presented in section \ref{sec:3spdc+3qubits-publication}. The purpose of that article is two-fold. On one hand, its section II introduces the theoretical concept of (non-)Gaussian entanglement and equips us with tools for proving the non-Gaussianity of some states. In particular, we use those tools to prove the non-Gaussianity of the \textit{GHZ} state, and the Gaussianity of the $W$ state was already proved in the literature \cite{teh_reid2019}, despite not being interpreted this way. On the other hand, we propose the design of a hybrid superconducting circuit with three pairs of bosonic modes and qubits that puts the newly introduced theoretical concepts to the test, in order to prove that they are adequate for actual experiments. In section III we prove the non-Gaussianity of the field's states produced from vacuum and 3SPDC amplification. Then, in section IV we prove the same non-Gaussianity on the qubit's reduced state, introducing a witness that is the DV counter part of the witness presented in chapter \ref{chapter1}. Then, the article's Conclussions and Appendices follow, which are advisable to read before continuing with the sections outside the article. 
	\section{Publication}
        \label{sec:3spdc+3qubits-publication}
        In this section the article \cite{agusti_sabin2022} is copied verbatim, as the main section of the present chapter. Please note it has its own page numbering as well as bibliography.
        \includepdf[pages=-]{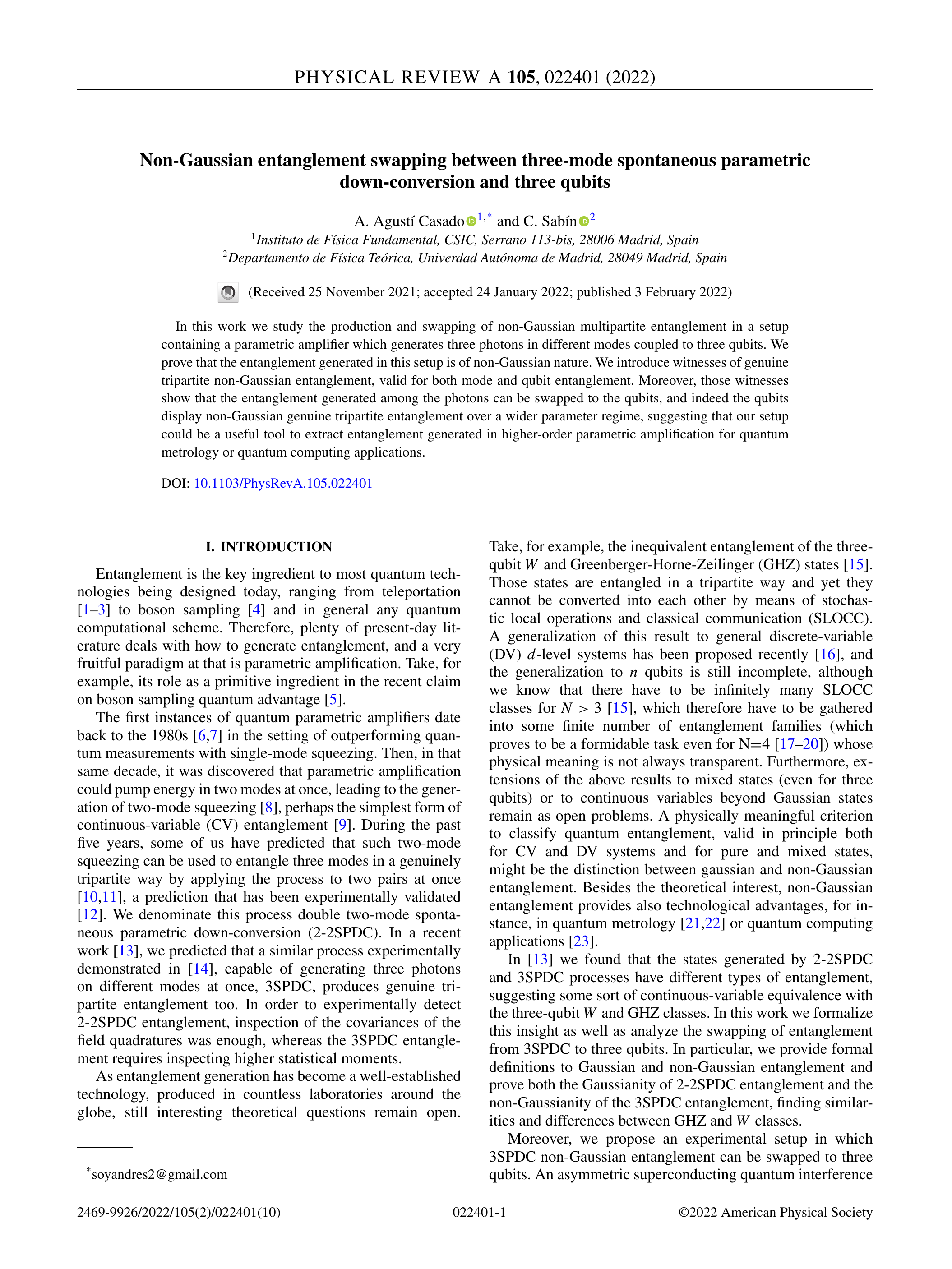}
    \section{Conclusions}
        This chapter \ref{chapter3}, that revolves around the results presented in \cite{agusti_sabin2022}, strengthens our results about the non-Gaussian nature of the entanglement in the 3SPDC system presented in chapter \ref{chapter1}. Additionally it generalizes our results to DV systems, displaying interesting analogies with the $W$ and $GHZ$ states, as well as proposing a device capable of producing non-Gaussian entanglement between qubits. That entanglement is reported in a wider parameter regime. Both facts could enhance the technological significance of non-Gaussian entanglement.
    \bibliographystyle{apsrev4-2}
    \bibliography{thesis.bib}{}

	\backmatter
\end{document}